\documentclass[11pt]{article}
\usepackage{fullpage}
\usepackage{setspace}
\usepackage{comment}
\usepackage[pdftex]{graphicx}
\usepackage{rotating}
\usepackage{amsmath}
\usepackage{amsthm}
\usepackage{amssymb}
\usepackage{graphicx}
\usepackage{fullpage}
\usepackage{setspace}
\usepackage{natbib}
\usepackage{color}
\usepackage{transparent}
\usepackage{multirow}
\usepackage{algorithm}
\usepackage[noend]{algpseudocode}
\usepackage[Symbolsmallscale]{upgreek}
\usepackage{centernot}
\algnewcommand{\LineComment}[1]{\State \(\triangleright\) #1}

\newcommand{\pn}{\mathbb{P}_n}

\newcommand{\T}{\intercal}

\newcommand{\bA}{ \mbox{\bf A}}
\newcommand{\ba}{\mathbf{a}}
\newcommand{\bB}{ \mbox{\bf B}}

\newcommand{\bY}{ \mbox{\bf Y}}

\newcommand{\bs}{ \mbox{\bf s}}

\newcommand{\bU}{ \mbox{\bf U}}

\newcommand{\bS}{\mathbf{S}}

\newcommand\independent{\protect\mathpalette{\protect\independenT}{\perp}}
\newcommand\dependent{\centernot{\independent}}
\def\independenT#1#2{\mathrel{\rlap{$#1#2$}\mkern2mu{#1#2}}}

\newtheorem{thm}{Theorem}[section]
\newtheorem*{thm*}{Theorem}

\newtheorem{lem}[thm]{Lemma}
\newtheorem{cor}[thm]{Corollary}
\theoremstyle{definition}
\newtheorem{defn}[thm]{Definition}

\begin{document}

\title{Sufficient Markov Decision Processes with Alternating Deep Neural Networks}
\author{Longshaokan Wang\\ Department of Statistics\\ North Carolina State University
\and Eric B. Laber \\ Department of Statistics\\ North Carolina State University
\and Katie Witkiewitz \\ Department of Psychology\\ University of New Mexico}


\pagestyle{empty}
\begin{center}
  \textbf{Sufficient Markov Decision Processes with Alternating Deep Neural Networks} \\
  \textbf{Longshaokan Wang$^{1}$, Eric B. Laber$^{1}$, Katie
    Witkiewitz$^{2}$}\\
  $^1$Department of Statistics, North Carolina State University, Raleigh, NC, 27695, U.S.A.  \\
  $^2$Department of Psychology, University of New Mexico, Albuquerque, NM, 87106, U.S.A. \\
\end{center} 
\begin{abstract}\noindent 
  Advances in mobile computing technologies have made it possible to
  monitor and apply data-driven interventions across complex systems
  in real time.  Recent and high-profile examples of data-driven
  decision making include autonomous vehicles, intelligent power
  grids, and precision medicine through mobile health.  Markov
  decision processes are the primary mathematical model for sequential
  decision problems with a large or indefinite time horizon; existing
  methods for estimation and inference rely critically on the
  correctness of this model.  Mathematically, this choice of model
  incurs little loss in generality as any decision process evolving in
  discrete time with observable process states, decisions, and
  outcomes can be represented as a Markov decision process.  However,
  in some application domains, e.g., mobile health, choosing a
  representation of the underlying decision process that is both
  Markov and low-dimensional is non-trivial; current practice is to
  select a representation using domain expertise.  We propose an
  automated method for constructing a low-dimensional representation
  of the original decision process for which: (P1) the Markov decision
  process model holds; and (P2) a decision strategy that leads to
  maximal mean utility when applied to the low-dimensional
  representation also leads to maximal mean utility when applied to
  population of interest.   Our approach uses a novel deep neural network
  to define a class of potential process representations and then
  searches within this class for the representation of lowest
  dimension which satisfies (P1) and (P2).  We illustrate the proposed
  method using a suite of simulation experiments and application to
  data from a mobile health intervention targeting smoking and heavy
  episodic drinking among college students.
\end{abstract} 

\pagebreak 
\setcounter{page}{1} 
\pagestyle{plain} 

\section{Introduction}
Sequential decision problems arise in a wide range of application
domains including autonomous vehicles \citep[][]{bagnell2001autonomous},
finance \citep[][]{bauerle2011markov}, logistics
\citep[][]{zhang1995reinforcement}, robotics
\citep[][]{kober2013reinforcement}, power grids
\citep[][]{riedmiller2000reinforcement}, and healthcare
\citep[][]{chakraborty2013statistical}.  Markov decision processes
(MDPs) \citep[][]{bellman1957markovian, puterman2014markov} are the
primary mathematical model for representing sequential decision
problems with an indefinite time horizon \citep[][]{Bertsekas1996,
  sutton1998, bather2000decision, si2004handbook,
  powell2007approximate, wiering2012reinforcement}.  This class of
models is quite general as almost any decision process can be made
into an MDP by concatenating data over multiple decision points (see
Section 2 for a precise statement); however, coercing a decision
process into the MDP framework in this way can lead to
high-dimensional system state information that is difficult to model
effectively.  One common approach to construct a low-dimensional
decision process from a high-dimensional MDP is to create a finite
discretization of the space of possible system states and to treat the
resultant process as a finite MDP \citep[][]{Gordon1995, Murao1997,
  sutton1998, Kamio2004, Whiteson2007}.  However, such discretization
can result in a significant loss of information and can be difficult to apply
when the system state information is continuous and high-dimensional.
Another common approach to dimension reduction is to construct a
low-dimensional summary of the underlying system states, e.g., by
applying principal components analysis \citep[][]{Jolliffe1986},
multidimensional scaling \citep[][]{Borg1997}, or by constructing a
local linear embedding \citep[][]{Roweis2000}.  These approaches can
identify a low-dimensional representation of the system state but, as
we shall demonstrate, they need not retain salient features for making
good decisions.

The preceding methods seek to construct a low-dimensional
representation of a high-dimensional MDP with the goal of using the
low-dimensional representation to estimate an optimal decision
strategy, i.e., one that leads to maximal mean utility when applied to
the original process; however, they offer no guarantee that the
resulting process is an MDP or that a decision strategy estimated
using data from the low-dimensional process will perform well when
applied to the original process.  We derive sufficient conditions
under which a low-dimensional representation is an MDP, and that an
optimal decision strategy for this low-dimensional representation is
optimal for the original process.  We develop a hypothesis test for
this sufficient condition based on the Brownian distance covariance
\citep[][]{ Szekely2007, Szekely2009} and use this test as the basis
for selecting a low-dimensional representation within 
a class of deep neural networks.
The proposed estimator can be viewed as a novel variant of deep neural
networks for feature construction in MDPs.

In Section \ref{sec:MDP}, we review the MDP model for 
sequential decision making and define an optimal
decision strategy. 
In Section
\ref{sec:sufficientDefn}, 
we derive  conditions
under which a low-dimensional representation of an MDP is sufficient
for estimating an optimal decision strategy for the original process.  
In Section \ref{sec:ADNN}, we develop a new deep learning algorithm that is designed to produce low-dimensional representation that satisfies the proposed sufficiency condition. In Section \ref{sec:experiments}, we evaluate the performance of the proposed method in a suite of simulated experiments.  
In Section \ref{sec:mobileHealth},  we illustrate the proposed method using
data from a study of a mobile health intervention targeting 
smoking and heavy episodic drinking among college students 
\citep[][]{Witkiewitz2014}.
A discussion of future work is given in Section \ref{sec:discussion}.

\section{Setup and Notation}
\label{sec:MDP}
We assume that the observed data are
$\left\lbrace
\left(
\bS_{i}^1, A_{i}^1, U_{i}^1, \bS_{i}^2, \ldots, A_i^T, U_i^T, \bS_i^{T+1}
\right)
\right\rbrace_{i=1}^{n}$ which comprise
$n$ independent and identically distributed copies of the
trajectory $(\bS^1, A^1,  U^1, \bS^2,\ldots, \allowbreak A^T, U^T, \bS^{T+1})$ where:
$T\in\mathbb{N}$ denotes the observation time; 
$\bS^t \in\mathbb{R}^{p_t}$ denotes a summary of information 
collected up to time $t=1,\ldots, T$;  $A^t\in\mathcal{A} = \left\lbrace 1,\ldots, K\right\rbrace$ denotes
the decision made at time $t=1,\ldots, T$; and 
$U^t = U^t(\bS^{t}, A^t, \bS^{t+1})$ is a real-valued deterministic 
function of $(\bS^t, A^t, \bS^{t+1})$
that 
quantifies 
the momentary ``goodness'' of being in state $\bS^t$, making decision
$A^t$, and subsequently transitioning to state $\bS^{t+1}$.  We 
assume throughout that $\sup_t|U^t| \le M$ with probability one
for some fixed constant $M$.  
In applications like
mobile health,  the observed data might be collected in a pilot
study with a preset time horizon $T$ \citep[][]{maahs2012outpatient,
Witkiewitz2014}; however, the intent is to use these data to estimate 
an intervention strategy that will maximize some measure of cumulative
utility when applied over an indefinite time horizon
\citep[][]{ertefaie2014constructing, liao2015sample, luckett2016estimating}.  
Thus, we assume that 
$(\bS^1, A^1, U^1, \bS^2,\ldots, A^T, U^T, \bS^{T+1})$ comprises the first
$T$ observations of the process
$(\bS^1, A^1, U^1, \bS^2,\ldots)$. 
Furthermore, we assume  (A0) that this infinite process is 
Markov and homogeneous in that it satisfies
\begin{equation}\label{transKernel}
P\left(
\bS^{t+1}\in\mathcal{G}^{t+1} \bigg| A^t, \bS^t, \ldots, A^1, \bS^1
\right) = 
P\left(
\bS^{t+1}\in\mathcal{G}^{t+1} \bigg| A^t, \bS^t
\right),
\end{equation}
for all (measurable) subsets
$\mathcal{G}^{t+1}\subseteq \mathrm{dom}\,\bS^{t+1}$ and
$t \in\mathbb{N}$ and that the probability measure in
(\ref{transKernel}) does not depend on $t$.  For any process
$(\bS^1, A^1, \bS^2,\ldots)$ one can define
$\widetilde{\bS}^t = (\bS^t, A^{t-1},\ldots, \bS^{t-m_t})$, where
$m_t$ is chosen so that process
$\left( \widetilde{\bS}^1, A^t, \widetilde{\bS}^t, \ldots\right)$
satisfies (A0); to see this, note that the result holds trivially for
$m_t=t-1$. Furthermore, by augmenting the state with a variable for
time, i.e., defining the new state at time $t$ to be
$(\widetilde{S}^t, t)$, one can ensure that the probability measure in
(A0) does not depend on $t$.  In practice, $m_t$ is typically chosen
to be a constant, as letting the dimension of the state grow with time
makes extrapolation beyond the observed time horizon, $T$, difficult.
Thus, hereafter we assume that the domain of the state is constant
over time, i.e.,
$\mathrm{dom}\,\bS^t = \mathcal{S}\subseteq \mathbb{R}^p$ for all
$t\in\mathbb{N}$.  Furthermore, we assume that the utility is
homogeneous in time, i.e., $U^t = U(\bS^t, A^t, \bS^{t+1})$ for all
$t\in\mathbb{N}$.

A decision strategy, $\pi:\mathcal{S}\rightarrow \mathcal{A}$,  is a map
from states to decisions so that, under $\pi$, a decision maker presented
with $\bS^t=\mathbf{s}^t$ at time $t$ will select decision $\pi(\mathbf{s}^t)$.  
We define an optimal decision strategy using the language of
potential outcomes \citep[][]{rubin1978bayesian}.   
We use an overline to denote history so that 
$\overline{\ba}^t = (a^1,\ldots, a^t)$ and $\overline{\bs}^t = 
(\bs^1,\ldots, \bs^t)$.
The set of 
potential outcomes is 
$\mathbf{O}^* = \left\lbrace \bS^{*t}(\overline{\ba}^{t-1})\right\rbrace_{t\ge 1}$ where
$\bS^{*t}(\overline{\ba}^{t-1})$ is the potential state 
under $\overline{\ba}^{t-1}$ and
we have defined $\bS^{*1}(\overline{\ba}^0) = \bS^1$. 
Thus, the potential utility at time $t$ under $\overline{\ba}^t$ is
$U\left\lbrace \bS^{*t}(\overline{\ba}^{t-1}), a^t,
  \bS^{*(t+1)}(\overline{\ba}^t)
\right\rbrace$.
The potential state under a decision
strategy, $\pi$, is
$\bS^{*t}(\pi) = \sum_{\overline{\ba}^{t-1}}\bS^{*t}(\overline{\ba}^{t-1})
\prod_{v=1}^{t-1}1_{\pi\{\bS^{*v}(\overline{\ba}^{v-1})\}=a^v}$, 
and the potential utility under $\pi$ is $U^{*t}(\pi) = U\left[
\bS^{*t}(\pi), \pi\left\lbrace \bS^{*t}(\pi)
\right\rbrace,
\bS^{*(t+1)}(\pi)
\right]$.  Define the discounted mean utility under a decision
strategy, $\pi$, as
\begin{equation*}
V(\pi) = \mathbb{E}\left\lbrace 
\sum_{t\ge 1}\gamma^{t-1}U^{*t}(\pi)
\right\rbrace,
\end{equation*}
where $\gamma \in (0,1)$ is a discount factor that balances the trade-off between
immediate and long-term utility.    Given a class of decision strategies,
$\Pi$, an optimal decision strategy, $\pi^{\mathrm{opt}}\in\Pi$, satisfies
$V(\pi^{\mathrm{opt}}) \ge V(\pi)$ for all $\pi \in \Pi$.  

Define $\mu^t(\ba^t;\overline{\bs}^t, \overline{\ba}^{t-1}) = P\left(A^t=a^t\big|
\overline{\bS}^t=\overline{\bs}^t, \overline{\bA}^{t-1}=\overline{\ba}^{t-1}\right)$.  
To characterize
$\pi^{\mathrm{opt}}$ in terms of the data-generating model, we make 
the following assumptions for all $t\in \mathbb{N}$:  
(C1)  consistency, $\bS^t = \bS^{*t}(\overline{\bA}^{t-1})$; 
(C2) positivity, there exists $\epsilon  >0$ such that 
$\mu^t(\ba^t;\overline{\bS}^t, \overline{\bA}^{t-1}) \ge \epsilon$ with probability one
   for all $a^t\in\mathcal{A}$; and
(C3) sequential ignorability, $\mathbf{O}^* \perp A^t \big| \overline{\bS}^t,
\overline{\bA}^{t-1}$.  
These assumptions are standard in data-driven decision making \citep[][]{robins2004optimal, schulte2014q}.  Assumptions (C2) and (C3) hold by design in a randomized trial \citep[][]{liao2015sample, klasnja2015microrandomized} but are not verifiable in the data for observational studies.  Under these
assumptions, the joint distribution of $\left\lbrace \bS^{*t}(\pi)\right\rbrace_{t=1}^T$
is non-parametrically identifiable under the data-generating model for
any decision strategy $\pi$ and time horizon $T$. In our application, these
assumptions will enable us to construct low-dimensional features of the 
state that retain all relevant information for estimating
$\pi^{\mathrm{opt}}$ without having to solve the original MDP as an
intermediate step.

\section{Sufficient Markov Decision Processes} 
\label{sec:sufficientDefn}
If the states $\bS^t$ are high-dimensional it can be difficult to
construct a high-quality estimator of the optimal decision strategy;
furthermore, in applications like mobile health, storage and
computational resources on the mobile device are limited, making it
desirable to store only as much information as is needed to inform
decision making.  For any map
$\phi:\mathcal{S}\rightarrow \mathbb{R}^q$ define
$\bS_{\phi}^t = \phi(\bS^t)$. 
 We say that $\phi$ induces a sufficient
MDP for $\pi^{\mathrm{opt}}$ if
$(\overline{\bA}^{t}, \overline{\bS}_{\phi}^{t+1}, \overline{\bU}^t)$
contains all relevant information in
$(\overline{\bA}^{t}, \overline{\bS}^{t+1}, \overline{\bU}^t)$ about
$\pi^{\mathrm{opt}}$. 
Given a policy $\pi_{\phi}:\mathrm{dom}\,\bS_{\phi}^t\rightarrow
\mathcal{A}$ define the potential utility under $\pi_{\phi}$ as
\begin{equation*}
U_{\phi}^{*t}(\pi_{\phi}) = 
\sum_{\overline{\ba}^t}U\left\lbrace
\bS^{*t}\left(\overline{\ba}^{t-1}\right), a^t, \bS^{*(t+1)}\left(\overline{\ba}^t\right)
\right\rbrace \prod_{v=1}^{t}1_{\pi_{\phi}\left\lbrace
    \bS_{\phi}^{*v}(\overline{\ba}^{v-1})
    \right\rbrace = a^v}.  
\end{equation*}
The following definition formalizes the
notion of  inducing a sufficient MDP.
\begin{defn} 
  Let $\Pi \subseteq \mathcal{A}^{\mathcal{S}}$ denote a class of
  decision strategies defined on $\mathcal{S}$ and
  $\Pi_{\phi}\subseteq \mathcal{A}^{\mathcal{S}_{\phi}}$ a class of
  decision strategies defined on
  $\mathcal{S}_{\phi}=\mathrm{dom}\,\bS_{\phi}^t\subseteq
  \mathbb{R}^q$.
  We say that the pair $(\phi, \Pi_{\phi})$ induces a sufficient MDP
  for $\pi^{\mathrm{opt}}$ within $\Pi$ if the following conditions
  hold for all $t\in\mathbb{N}$:
\begin{itemize}
  \item[(SM1)]  the process $(\overline{\bA}^{t}, \overline{\bS}_{\phi}^{t+1}, 
\overline{\bU}^t)$ is Markov and homogeneous, i.e., 
\begin{equation*}
P\left(
\bS_{\phi}^{t+1} \in \mathcal{G}_{\phi}^{t+1}\big|
\overline{\bS}_{\phi}^{t}, \overline{\bA}^{t}
\right) = 
P\left(
\bS_{\phi}^{t+1} \in \mathcal{G}_{\phi}^{t+1}\big|
{\bS}_{\phi}^{t}, {A}^{t}
\right)
\end{equation*}
for any (measurable) subset $\mathcal{G}_{\phi}^{t+1}\subseteq \mathbb{R}^q$
and this probability does not depend on $t$;
\item[(SM2)] there exists $\pi^{\mathrm{opt}} \in
  \arg\max_{\pi\in\Pi}V(\pi)$ which can be 
  written as $\pi^{\mathrm{opt}} = \pi_{\phi}^{\mathrm{opt}}\circ
  \phi$, where
  $\pi_{\phi}^{\mathrm{opt}} \in
  \arg\max_{\pi_{\phi}\in\Pi_{\phi}}\mathbb{E}
  \left\lbrace
  \sum_{t\ge 1}
  \gamma^{t-1}U_{\phi}^{*t}(\pi_{\phi})\right\rbrace$.  
\end{itemize}
\end{defn}
\noindent
Thus, given observed data, $\left\lbrace (\overline{\bS}_{i}^{T+1},
 \overline{\bA}_i^{T}, \overline{\bU}_i^{T})\right\rbrace_{i=1}^{n}$
and class of decision strategies, $\Pi$, if one can 
find a pair $(\phi, \Pi_{\phi})$ which induces a sufficient
MDP for $\pi^{\mathrm{opt}}$ within $\Pi$, then it suffices
to store only the reduced process $\left\lbrace \left(\overline{\bS}_{\phi,i}^{T+1},
\overline{\bA}_{i}^{T}, \overline{\bU}_{i}^{T}\right)\right\rbrace_{i=1}^{n}$.
Furthermore, existing reinforcement learning algorithms
\citep[e.g.,][]{sutton1998, szepesvari2010algorithms}  can be applied
to this reduced process to construct an estimator of $\pi_{\phi}^{\mathrm{opt}}$ 
and hence $\pi^{\mathrm{opt}} = \pi_{\phi}^{\mathrm{opt}}\circ\phi$.  
If the dimension of $\bS_{\phi}^t$ is substantially smaller than 
that of $\bS^t$, then using the reduced process can lead to smaller estimation 
error as well as reduced storage and computational costs.  In some
applications, it may also be desirable to have $\phi$ be a sparse function
of $\bS^t$ in the sense that it only depends on a subset of the
components of $\bS^t$. For example, in the context of mobile health,
one may construct the state, $\bS^t$, by concatenating measurements
taken at time points $t, t-1,\ldots, t-m$, where the look-back period,
$m$, is chosen conservatively based on clinical judgement to ensure
that 
the process is Markov; however, a data-driven sparse feature map might
identify that a look-back period of $m' \ll m$ is sufficient thereby
reducing
computational and memory requirements but also generating new
knowledge that may be of clinical value.  
The remainder of this
section will focus on developing verifiable conditions for checking that
$(\phi, \Pi_{\phi})$ induces a sufficient MDP.  These conditions
are used to build a data-driven, low-dimensional, and potentially sparse
sufficient MDP.

Define $\bY^{t+1} = \left\lbrace U^t, (\bS^{t+1})^{\T}\right\rbrace^{\T}$
for all $t\in\mathbb{N}$. 
The following result provides a conditional independence criterion 
that ensures a given feature
map induces a sufficient MDP; this criterion can be seen
as an MDP analog of 
nonlinear sufficient dimension reduction in 
regression \citep[][]{cook2007fisher, li2011principal}.  
A proof is provided in the Supplemental Materials.
\begin{thm}\label{theoremey}
Let $(\bS^1, A^1, U^1, \bS^2,\ldots)$ be an MDP that satisfies (A0)
and (C1)-(C3).
 Suppose that there exists $\phi:\mathcal{S}\rightarrow
\mathbb{R}^q$ such that
\begin{equation}\label{sufficientTheoremCond1}
\bY^{t+1}\independent \bS^{t} \big| \bS_{\phi}^t, A^t,
\end{equation}
then, $(\phi, \Pi_{\phi,\mathrm{msbl}})$ induces a sufficient MDP for 
$\pi^{\mathrm{opt}}$ within $\Pi_{\mathrm{msbl}}$, where
$\Pi_{\mathrm{msbl}}$ is the set of measurable maps from $\mathcal{S}$
into $\mathcal{A}$ and $\Pi_{\phi,\mathrm{msbl}}$ is the set of measurable
maps from $\mathbb{R}^q$ into $\mathcal{A}$.  
\end{thm}
\noindent
The preceding result could be used to construct an estimator for
$\phi$ so that $(\phi,\Pi_{\phi,\mathrm{msbl}})$ induces a sufficient
MDP for $\pi^{\mathrm{opt}}$ within $\Pi_{\mathrm{msbl}}$ as follows.
Let $\Phi$ denote a potential class of vector-valued functions on
$\mathcal{S}$.  Let $\widehat{p}_{n}(\phi)$ denote a p-value for
a test of the conditional independence criterion
(\ref{sufficientTheoremCond1}) based on the mapping $\phi$, e.g., one
might construct this p-value using conditional Brownian distance
correlation \citep[][]{Wang2015} or kernel-based tests of conditional
independence \citep[][]{fukumizu2007kernel}.  Then, one could select
$\widehat{\phi}_{n}$ to be the transformation of
lowest dimension among those within the set
$\left\lbrace \phi\in\Phi\,:\, \widehat{d}_{n}(\phi) \ge
  \tau\right\rbrace$,
where $\tau$ is a fixed significance level, e.g., $\tau=0.10$.
However, such an approach can be computationally burdensome especially
if the class $\Phi$ is large.  Instead, we will develop a procedure
based on a series of unconditional tests that is computationally simpler
and allows for a flexible class of potential transformations.  Before
presenting this approach, we first describe how the conditional
independence criterion in the above theorem can be applied recursively
to potentially produce a sufficient MDP of lower dimension.

The condition $\bY^{t+1}\independent \bS^{t} \big| \bS_{\phi}^t, A^t$
is overly stringent in that it requires $\bS_{\phi}^t$ to capture all
the information about $\bY^{t+1}$ contained within $\bS^t$ regardless of
whether or not that information is useful for decision making.
However, given a sufficient MDP
$(\bS_{\phi}^1, A^1, U^1, \bS_{\phi}^{2},\ldots)$, 
one can apply the above theorem to this
MDP to obtain further dimension reduction; this process can 
be iterated until no further dimension reduction is possible. 
For any map $\phi:\mathcal{S}\rightarrow \mathbb{R}^q$, define
 $\bY_{\phi}^t = \left\lbrace U^{t},
  \left(\bS_{\phi}^{t+1}\right)^{\T}
\right\rbrace^{\T}$.
The following result is proved in the Supplemental Materials.  
\begin{cor}\label{coraline}
  Let $(\bS^1, A^1, U^1, \bS^2,\ldots)$ be an MDP that satisfies (A0) and
  (C1)-(C3).  
  Assume that there exists $\phi_0:\mathcal{S}\rightarrow
  \mathbb{R}^{q_0}$ such that $(\phi_{0}, \Pi_{\phi_0,
    \mathrm{msrbl}})$  induces a sufficient MDP for
  $\pi^{\mathrm{opt}}$ 
  within $\Pi_{\mathrm{msrbl}}$.    Suppose that there exists
  $\phi_1:\mathbb{R}^{q_0}\rightarrow\mathbb{R}^{q_1}$ such that for
  all $t\in\mathbb{N}$ 
  \begin{equation}\label{sufficientCorCond1}
    \bY_{\phi_0}^{t+1}\independent \bS_{\phi_0}^{t}\big|
    \bS_{\phi_1\circ\phi_0}^{t}, A^t,
  \end{equation} 
  then $(\phi_1\circ\phi_0, \Pi_{\phi_1\circ\phi_0,\mathrm{msrbl}})$ induces a sufficient
  MDP for $\pi^{\mathrm{opt}}$ within $\Pi_{\mathrm{msrbl}}$.
  Furthermore, for $k\ge 2$, denoting $\phi_{k}\circ\phi_{k-1}\circ\cdots\circ \phi_0$ as $\overline{\phi}_k$, if there exists
  $\phi_{k}:\mathbb{R}^{q_{k-1}}\rightarrow \mathbb{R}^{q_k}$ such
  that
  $\bY_{\overline{\phi}_{k-1}}^{t+1}\independent
  \bS_{\overline{\phi}_{k-1}}^{t}\big|\bS_{\overline{\phi}_{k}}^{t}, A^t$, 
  then $(\overline{\phi}_{k},
  \Pi_{\overline{\phi}_{k},
\mathrm{msrbl}})$ induces a sufficient
  MDP for $\pi^{\mathrm{opt}}$ within $\Pi_{\mathrm{msrbl}}$.
\end{cor}
\noindent
We now state a simple condition involving the residuals of a
multivariate regression that can be used to test the conditional
independence required in each step of the preceding corollary.  In our
implemenation we use residuals from a varient of deep neural networks
that is suited to sequential decision problems (see Section 4).
The following result is proved in the Supplemental Materials.  
\begin{lem}\label{llama}
Let $(\bS^1, A^1, U^1, \bS^2,\ldots)$ be an MDP that satisfies (A0) and
(C1)-(C3).  Suppose that there exists $\phi:\mathcal{S}\rightarrow
\mathbb{R}^q$ such that at least one of the following conditions hold:
\begin{itemize}
  \item[(i)]  $\left\lbrace \bY^{t+1} - \mathbb{E}\left(
        \bY^{t+1}\big|\bS_{\phi}^{t}, A^t\right)       \right\rbrace \independent 
      \bS^t
     \big| A^t$,
  \item[(ii)] $\left\lbrace
      \bS^t-\mathbb{E}\left(
        \bS^t\big|\bS_{\phi}^t
        \right)
      \right\rbrace \independent
        \left(\bY^{t+1}, \bS_{\phi}^t\right)
         \big| A^t,
      $
\end{itemize}
then  $\bY^{t+1}\independent \bS^t \big| \bS_{\phi}^t, A^t$.  
\end{lem}
\noindent 
The preceding result can be used to verify the conditional
independence condition required by Theorem (\ref{theoremey}) and
Corollary (\ref{coraline}) 
using unconditional tests of independence within levels of
$A^t$; in our simulation experiments, we used Brownian
distance covariance for continuous states \citep[][]{Szekely2007,
    Szekely2009}
and a likelihood ratio test for discrete states, though other 
choices are possible \citep[][]{gretton2005measuring, gretton2005kernel}.    Application of these tests requires
modification to account for dependence over time
within each subject.  One simple approach, the one we follow
here, is to compute a separate test at each time point and then
to pool the resultant p-values using a pooling procedure
that allows for general dependence.  For example, let
$G^t = g(\bS^t, A^t, \bS^{t+1})\in\mathbb{R}^{d_1}$ and 
$H^t=h(\bS^t)\in\mathbb{R}^{d_2}$ be known features
of $(\bS^t, A^t, \bS^{t+1})$  and $\bS^t$.   Let $\pn$ denote
the empirical measure.  To test 
$G^t \independent H^t$ using the Brownian distance covariance,
we compute the test statistic 
\begin{eqnarray*}
 \widehat{\mathbb{T}}_n^t &=& ||\pn \exp\left\lbrace
\mathrm{i}\left(
\varsigma^{\T}G^t + \varrho^{\T}H_t
\right)
\right\rbrace
- 
\pn \exp\left(
\mathrm{i}
\varsigma^{\T}G^t
\right)
\pn \exp\left(
\mathrm{i}
\varrho^{\T}G^t
\right)
||_{\omega}^2  \\ & =& 
\int \frac{
\left[\pn \exp\left\lbrace
\mathrm{i}\left(
\varsigma^{\T}G^t + \varrho^{\T}H_t
\right)
\right\rbrace
- 
\pn \exp\left(
\mathrm{i}
\varsigma^{\T}G^t
\right)
\pn \exp\left(
\mathrm{i}
\varrho^{\T}H^t
\right)\right]^2 
\Gamma\left(\frac{
1+d_1
}{
2
}
\right)
\Gamma\left(
\frac{
1+d_2
}{
2
}
\right)
}
{
||\varsigma||^{d_1+1}||\varrho||^{d_2+1}
\uppi^{(d_1+d_2+2)/2}}d\varrho d\varsigma,
\end{eqnarray*}
and subsequently compute the $p$-value, say $\widehat{p}_n^t$, using
the null distribution of $\widehat{\mathbb{T}}_{n}$ as estimated with permutation \citep[see][for details]{Szekely2007,
    Szekely2009}. For each $u=1,\ldots, T$,
 let $\widehat{p}_{n}^{(u)}$ denote the $u$th order
  statistic of $\widehat{p}_{n}^{1},\ldots, \widehat{p}_{n}^T$ and 
define the pooled $p$-value 
$\widehat{p}_{n,\mathrm{pooled}}^{u} = T\widehat{p}_{n}^{(u)}/u$.  
For each $u=1,\ldots, T$ it can be shown that $\widehat{p}_{n,\mathrm{pooled}}^{u}$ is valid
$p$-value \citep[][]{ruger1978das}, e.g., $u=1$ corresponds to the
common Bonferroni correction.  In our simulation experiments, we
set $u = \lfloor T/20 + 1 \rfloor$ across all settings.

\subsection{Variable screening}
The preceding results provide a pathway for constructing sufficient
MDPs.  However, while the criteria given in Theorem
\ref{theoremey} and Lemma \ref{llama} can be used to identify
low-dimensional structure in the state, they cannot be used to
eliminate certain
simple types of noise variables.  For example, let
$\left\lbrace \bB^t\right\rbrace_{t\ge 1}$ denote a homogeneous Markov
process that is independent of $(\bS^1, A^1, U^1,\bS^2\ldots)$, and
consider the augmented process
$\left(\widetilde{\bS}^1, A^1, U^1, \widetilde{\bS}^2, \ldots\right)$,
where
$\widetilde{\bS}^t = \left\lbrace (\bS^t)^{\T},
  (\bB^t)^{\T}\right\rbrace^{\T}$.
Clearly, the optimal policy for the augmented process does not depend
on $\left\lbrace \bB^t \right\rbrace_{t\ge 1}$, yet, $\bY^{t+1}$ need
not be conditionally independent of $\widetilde{\bS}^t$ given
$\bS^t$. To remove variables of this type, we develop a simple
screening procedure that can be applied prior to constructing
nonlinear features as described in the next section.

The proposed screening procedure is based on the following
result which is proved in the Supplemental Materials.
\begin{thm}\label{strongThm}
Let $(\bS^1, A^1, U^1, \bS^2,\ldots)$ be an MDP that satisfies (A0)
and (C1)-(C3).
 Suppose that there exists $\phi:\mathcal{S}\rightarrow
\mathbb{R}^q$ such that
\begin{equation}\label{sufficientTheoremCond2}
\bY_{\phi}^{t+1}\independent \bS^{t} \big| \bS_{\phi}^t, A^t,
\end{equation}
then, $(\phi, \Pi_{\phi,\mathrm{msbl}})$ induces a sufficient MDP for 
$\pi^{\mathrm{opt}}$ within $\Pi_{\mathrm{msbl}}$, where
$\Pi_{\mathrm{msbl}}$ is the set of measurable maps from $\mathcal{S}$
into $\mathcal{A}$ and $\Pi_{\phi,\mathrm{msbl}}$ is the set of measurable
maps from $\mathbb{R}^q$ into $\mathcal{A}$.  
\end{thm}\noindent
This result can be viewed as a stronger version of Theorem
\ref{theoremey} in that the required conditional independence
condition is weaker; indeed, in the example stated above, it can be
seen that $\phi(\widetilde{\bS}^t) = \bS^t$ satisfies
(\ref{sufficientTheoremCond2}).  However, because $\phi$ appears in
both $\bY_{\phi}^{t+1}$ and $\bS_{\phi}^t$, constructing nonlinear
features using this criterion is more challenging as the
residual-based conditions stated in Lemma \ref{llama} can no longer be
applied.  Nevertheless, this criterion turns out to be ideally suited
to screening procedures wherein the functions
$\phi:\mathbb{R}^p\rightarrow
\mathbb{R}^q$ are of the form
$\phi(\bs^t)_j =s_{k_j}^t$ for $j=1,\ldots, q$, where $\left\lbrace k_1,\ldots,
k_q\right\rbrace$ is a subset of $\left\lbrace 1,\ldots, p
\right\rbrace$.   

For any subset $J \subseteq \left\lbrace 1,\ldots, p\right\rbrace$, 
define $\bS_{J}^t = \left\lbrace S_j^t\right\rbrace_{j\in J}$ and
$\bY_J^t = \left\lbrace U^t, (\bS_J^t)^{\T}\right\rbrace$.   
Let $J_1$ denote the smallest set of indices such that 
$U^t$ depends on $\bS^t$ and $\bS^{t+1}$ only through
$\bS_{J_1}^{t}$ and $\bS_{J_1}^{t+1}$ conditioned on $A^t$.   For $k \ge 2$, define 
$J_k = \left\lbrace 1\le j \le p\,:\, S_j^t \dependent 
\bY_{J_{k-1}}^t \big| A^t \right\rbrace$. Let $K$ denote the smallest value
for which
$J_{K-1} = J_{K}$, such a $K$ must exist as $J_{k-1} \subseteq J_{k}$
for all $k$, and define $\phi_{\mathrm{screen}}(\bS^t) =
\bS_{J_K}^{t}$.    The following results shows that
$\phi_{\mathrm{screen}}$ induces a sufficient MDP; furthermore, 
Corollary \ref{sufficientCorCond1} shows that such screening
can be applied before nonlinear feature construction without
destroying sufficiency.
\begin{thm}\label{screenThm}
Let $(\bS^1, A^1, U^1, \bS^2,\ldots)$ be an MDP that satisfies (A0)
and (C1)-(C3), and let $J_1,\ldots, J_K, \phi_{\mathrm{screen}}$ be as
defined above.
Assume that for any two non-empty subsets, $J, J' \subseteq
\left\lbrace 1,\ldots, p\right\rbrace$, if 
$\bS_J^t \dependent \bY_{J'}^{t+1} \big| A^t$ 
then there exists $j\in J$ such  that $S_j^t \dependent
\bY_{J'}^{t+1} \big| A^t$.  Then, $\bY_{\phi_{\mathrm{screen}}}^{t+1}
\independent \bS^t \big| \bS_{\phi_{\mathrm{screen}}}^{t}, A^t$.  
\end{thm}\noindent
The condition that joint dependence implies marginal
dependence (or equivalently, marginal independence implies joint independence) ensures that screening one variable at a
time will identify the entire collection of important
variables; this condition could be weakened by considering
sets of multiple variables at a time though at the expense of additional
computational burden.  Algorithm \ref{screenAlg} gives a schematic
for estimating $\phi_{\mathrm{screen}}$ using the Brownian distance
covariance to test for dependence.   The inner for-loop (lines 4-7)
of the algorithm can be executed in parallel and thereby scaled
to large domains.

\begin{algorithm}
\caption{\label{screenAlg}Screening with Brownian Distance Covariance}
\begin{algorithmic}[1]
\renewcommand{\algorithmicrequire}{\textbf{Input:}}
\renewcommand{\algorithmicensure}{\textbf{Output:}}
\algnewcommand\algorithmicforeach{\textbf{for each}}
\algdef{S}[FOR]{ForEach}[1]{\algorithmicforeach\ #1\ \algorithmicdo}
\Require{p-value threshold $\tau$; max number of iterations $N_{\max}$; data $\left\lbrace (\overline{\bS}_i^{T+1}, \overline{\bA}_i^T, \overline{\mathbf{U}}_i^T)\right\rbrace_{i=1}^{n}$; set of all indices $D = \{1, 2, \ldots, p = \text{dim}(\bS^t)\}$.}
\State Set $J_0 = \emptyset$, and $\mathbf{Y}^{t+1}_{J_0} = \{U^t\}$
\For{$k=1,\ldots,N_{\max}$}
\State Set $J_k = J_{k-1}$
\ForEach{$j \in D \setminus J_{k-1}$}
\State Perform dCov test on $S_j^t$ and $\mathbf{Y}^{t+1}_{J_{k-1}}$ within levels of $A^t$
\If{p-value $\leq \tau$}
\State Set $J_k = J_k \cup \{j\}$
\EndIf
\EndFor
\If{$J_k = J_{k-1}$}
\State Set $K = k$, stop.
\EndIf
\EndFor
\Ensure{$J_K$}
\end{algorithmic}
\end{algorithm}

\section{Alternating Deep Neural Networks}
\label{sec:ADNN}
For simplicity, we assume that $\mathcal{S} = \mathbb{R}^p$.  We
consider summary functions $\phi:\mathbb{R}^p\rightarrow \mathbb{R}^q$
that are representable as multi-layer neural networks
\citep[][]{anthony2009neural, Lecun2015, bengio2016deep}.  Multi-layer
neural networks have recently become a focal point in machine learning
research because of their ability to identify complex and nonlinear
structure in high-dimensional data \citep[see][and references
therein]{bengio2016deep}.  Thus, such models are ideally suited for
nonlinear feature construction; here, we present a novel neural
network architecture for estimating sufficient MDPs.

\begin{figure}[!h]
	\makebox[\textwidth][c]{\includegraphics[width=0.85\textwidth]{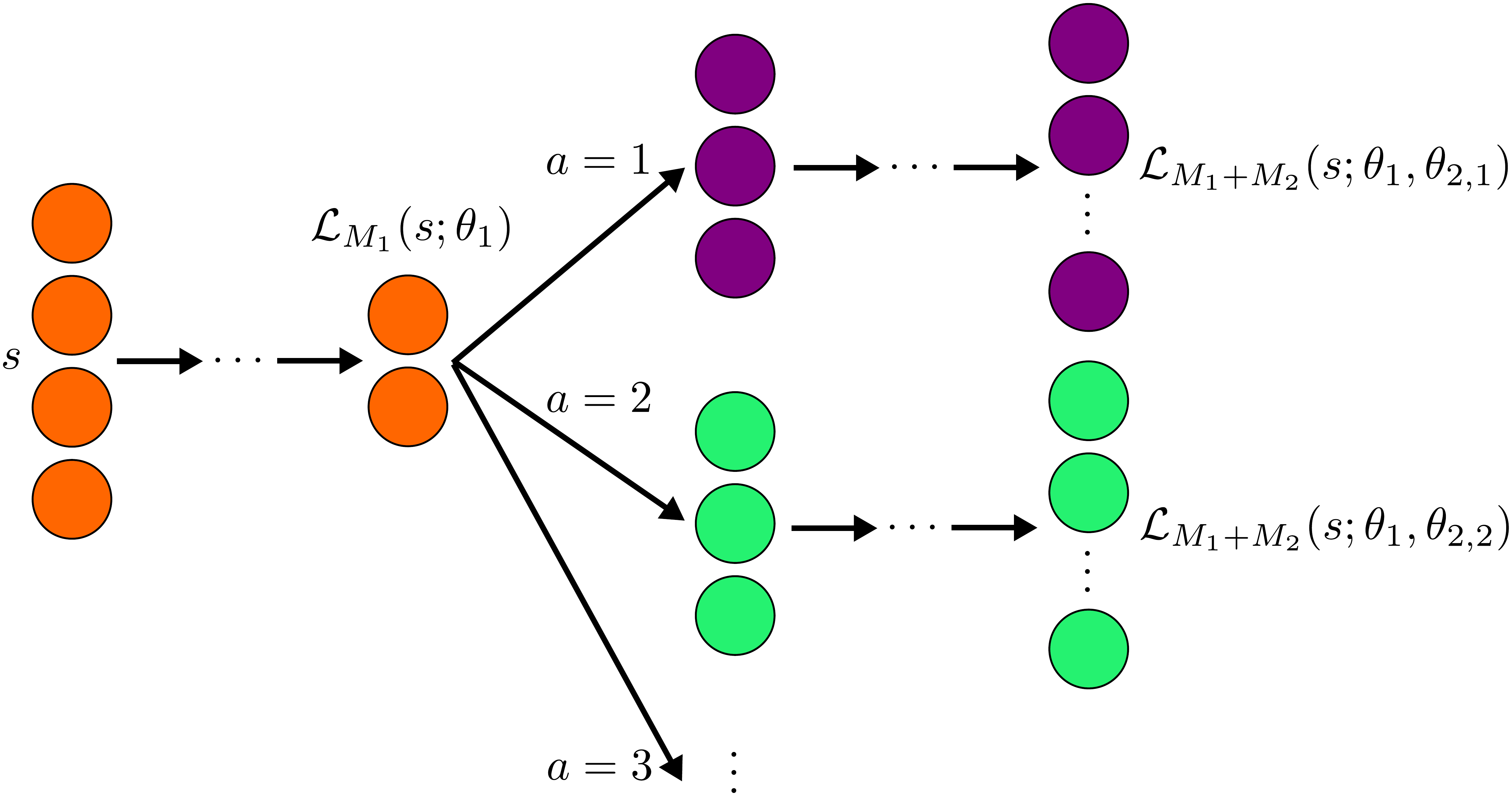}}
	\caption{\label{adnnSchematic}
          Schematic for alternating deep neural network (ADNN) model.  The
          term `alternating' refers to the estimation algorithm which
          cycles
          over the networks for each treatment $a\in\mathcal{A}$. 
        }
\end{figure}

We use criteria (i) in Lemma (\ref{llama}) to construct a data-driven
summary function $\phi$, therefore we also require a model for the
regression of $\bY^{t+1}$ on $\bS_{\phi}^t$ and $A^t$; we also use a
multi-layer neural network for this predictive model.  Thus, the model
can be visualized as two multi-layer neural networks: one that
composes the feature map $\phi$ and another that models the regression
of $\bY^{t+1}$ on $\bS_{\phi}^t$ and $A^t$.  A schematic for this
model is displayed in Figure \ref{adnnSchematic}.  Let
$\Phi:\mathbb{R}\rightarrow [0,1]$ denote a continuous and monotone
increasing function and write $\Phi^{\circ}$ to denote the
vector-valued 
function obtained by elementwise application of $\Phi$, i.e.,
$\Phi_j^{\circ}(v) = \Phi(v_j)$ where $v\in\mathbb{R}^d$.  The neural
network for the feature map is parameterized as follows.  Let
$r_1,\ldots, r_{M_1}\in\mathbb{N}$ be such that $r_1 = p$. The first
layer of the feature map network is
$\mathcal{L}_1(\bs; \Sigma_1, \eta_1) = \Phi^{\circ}\left( \Sigma_1\bs
  + \eta_1 \right)$,
where $\Sigma_1 \in \mathbb{R}^{r_2\times r_1}$ and
$\eta_1 \in \mathbb{R}^{r_2}$.  Recursively, for $k=2,\ldots, M_1,$
define
\begin{equation*}
\mathcal{L}_{k}(\bs;\Sigma_k, \eta_k,\ldots, \Sigma_1, \eta_1) = 
\Phi^{\circ}\left\lbrace
\Sigma_k\mathcal{L}_{k-1}\left(\bs; \Sigma_{k-1}, \eta_{k-1}, \ldots,
\Sigma_{1}, \eta_{1}\right) 
+ \eta_{k}
\right\rbrace,
\end{equation*}
where $\Sigma_k\in\mathbb{R}^{r_{k}\times r_{k-1}}$ and
$\eta_{k}\in\mathbb{R}^{k}$.  Let $\theta_1 = 
\left(
\Sigma_{M_1}, \eta_{M_1}, \ldots, \Sigma_{1}, \eta_1
\right)$ then the feature map under $\theta$ is 
$\phi(\bs;\theta_1)  = \mathcal{L}_{M_1}(\bs; \theta_1)  = 
\mathcal{L}_{M_1}\left(\bs; \Sigma_{M_1}, \eta_{M_1},\ldots, \Sigma_1,
  \eta_1
\right)$.  Thus, the dimension of the feature map
is $r_{M_1}$.   
The neural network for the regression of $\bY^{t+1}$ 
on $\bS_{\phi}^t$ and $A^t$ is as follows.  Let $r_{M_1+1}, \ldots,
r_{M_1+M_2} \in \mathbb{N}$ be such that $r_{M_1+M_2} = p+1$. 
For each $a\in\mathcal{A}$ define
$\mathcal{L}_{M_1+1,a}(\bs; \theta_1, \Sigma_{M_1+1,a}, \eta_{M_1+1,a})  = 
\Phi^{\circ}\left\lbrace
\Sigma_{M_1+1,a}\phi(\bs;\theta_1) + \eta_{M_1,a}
\right\rbrace$, where $\Sigma_{M_1+1,a} \in
\mathbb{R}^{r_{M_1+1}\times r_{M_1}}$ and $\eta_{M_1,
  a}\in\mathbb{R}^{r_{M_1+1}}$.  
Recursively, for 
$k=2,\ldots, M_2$ and each $a\in\mathcal{A}$  
define 
\begin{multline*}
\mathcal{L}_{M_1+k,a}\left(\bs; \theta_1, \Sigma_{M_1+k,a},
\eta_{M_1+k,a},\ldots, \Sigma_{M_1+1,a}, \eta_{M_1+1,a}\right)
\\ = \Phi^{\circ}\left\lbrace
\Sigma_{M_1+k,a}\mathcal{L}_{M_1+k-1,a}\left(\bs; \theta_1,\Sigma_{M_1+k-1,a}, \eta_{M_1+k-1,a}, \ldots,
\Sigma_{M_1+1,a}, \eta_{M_1+1,a}\right) 
+ \eta_{M_1+k,a}
\right\rbrace,
\end{multline*}
where $\Sigma_{M_1+k,a}\in\mathbb{R}^{r_{M_1+k}\times r_{M_1+k-1}}$
and $\eta_{M_1+k}\in\mathbb{R}^{r_{M_1+k}}$.  
For each $a\in\mathcal{A}$, define
$\theta_{2,a} = \\ 
\left(\Sigma_{M_1+M_2,a}, \\
  \eta_{M_1+M_2,a},\allowbreak \ldots, \allowbreak
\Sigma_{M_1+1, a},\eta_{M_1+1,a}\right)$, and write $\theta_2 =
\left\lbrace \theta_{2,a}\right\rbrace_{a\in\mathcal{A}}$.  The
postulated
model for $\mathbb{E}\left(\bY^{t+1}\big|\bS_{\phi}^t = \bs_{\phi}^t,
  A^t=a^t\right)$ under 
parameters $(\theta_1, \theta_2)$ is 
$\mathcal{L}_{M_1+M_1}(\bs;\theta_1, \theta_{2,a^t})$.  

We use penalized least squares to construct an estimator of
$(\theta_1, \theta_2)$.  Let $\pn$ denote the empirical measure and
define
\begin{equation*}
C_{n}^{\lambda}(\theta_1, \theta_2) = \pn \sum_{t=1}^{T}||
\mathcal{L}_{M_1+M_2}(\bS^{t};\theta_1, \theta_{2,A^t}) - \bY^{t+1}
||^2 + \lambda \sum_{j=1}^{r_1}\sqrt{
\sum_{\ell=1}^{r_2}\Sigma_{1,\ell,j}^2
},
\end{equation*}
and subsequently
$\left(\widehat{\theta}_{1,n}^{\lambda},
  \widehat{\theta}_{2,n}^{\lambda}\right) = \arg\min_{\theta_1,
  \theta_2}C_n^{\lambda}(\theta_1, \theta_2)$,
where $\lambda > 0$ is a tuning parameter.  The term
$\sqrt{ \sum_{\ell=1}^{r_2}\Sigma_{1,\ell,j}^2}$ is a group-lasso
penalty \citep[][]{yuan2006model} on the $\ell$th column of
$\Sigma_1$; if the $\ell$th column of $\Sigma_1$ shrunk to zero then
$\bS_{\phi}^t$ does not depend on the $\ell$th component of $\bS^t$.
Computation of
$\left(\widehat{\theta}_{1,n}^{\lambda},
  \widehat{\theta}_{2,n}^{\lambda}\right)$
also requires choosing values for $\lambda, M_1, M_2$, and
$r_{2},\ldots, r_{M_1-1}, r_{M_1+1},\ldots, r_{M_1+M_2-1}$, (recall
that $r_1 = p$, $r_{M_1+M_2} = p+1$, and $r_{M_1}$ is the dimension of the
feature map and is therefore considered separately).  Tuning each of
these parameters individually can be computationally burdensome,
especially when $M_1+ M_2$ is large.  In our implementation, we
assumed
$r_2 = r_3 = \cdots = r_{M_1-1}=r_{M_1+1}=\cdots= r_{M_1+M_2-1}=K_1$
and $M_1=M_2 =K_2$; then, for each fixed value of $r_{M_1}$ we
selected $(K_1, K_2, \lambda)$ to minimize cross-validated cost.
Algorithm \ref{adnnAlg} shows the process for fitting this model; the
algorithm uses subsampling 
to improve stability of the underlying
sub-gradient descent updates \citep[this is also known as taking
minibatches, see][and references
therein]{tutorial2014lisa, bengio2016deep}.

\begin{algorithm}
\caption{\label{adnnAlg}Alternating Deep Neural Networks}
\begin{algorithmic}[1]
\renewcommand{\algorithmicrequire}{\textbf{Input:}}
\renewcommand{\algorithmicensure}{\textbf{Output:}}
\algnewcommand\algorithmicforeach{\textbf{for each}}
\algdef{S}[FOR]{ForEach}[1]{\algorithmicforeach\ #1\ \algorithmicdo}
\Require{Tuning parameters $K_1, K_2\in\mathbb{N}$, $\lambda \ge 0$;
  feature map dimension $r_1$; 
  data  $\left\lbrace
    \overline{\bS}_i^{T+1},
  \overline{\bA}_{i}^T, \overline{\mathbf{U}}^T_i \right\rbrace_{i=1}^{n}$; batch size proportion
$\nu \in (0,1)$;  gradient-descent step-size $\left\lbrace
  \alpha_{b}\right\rbrace_{b\ge 1}$; error tolerance $\epsilon > 0$;
max number of iterations $N_{\max}$; and initial parameter values 
$\widehat{\theta}_{1,n}^{(1)}, \widehat{\theta}_{2,n}^{(1)}$.}
\State Set $D_{a} = \left\lbrace (i, t) \,:\, A_i^t = a\right\rbrace$ and
$n_a = \#D_a$ for
each $a\in\mathcal{A}$ and $t=1,\ldots, T$
\For{$b=1,\ldots,N_{\max}$}
\ForEach{$a \in \mathcal{A}$}
\State Draw a random batch $B_a$ of size $\lfloor \nu n_a \rfloor$
without replacement  from $D_a$
\State Compute a sub-gradient of the cost on batch $B_a$
\begin{equation*}
\Lambda_{a}^{(b)} =
\nabla
\left[
\frac{1}{\lfloor \nu n_a\rfloor}
\sum_{(i,t) \in B_a}
\big|\big|
\mathcal{L}_{M_1+M_2}\left\lbrace
\bS_i^{t}; \widehat{\theta}_{1,n}^{(b)}, \widehat{\theta}_{2,a,n}^{(b)}
\right\rbrace - \bY_{i}^{t+1}
\big|\big|^2 + 
\lambda\sum_{j=1}^{r_1}\sqrt{\sum_{\ell=1}^{r_2}\Sigma_{1,\ell,j}^2}
\right] 
\end{equation*}
\State Compute a sub-gradient descent update 
\begin{equation*}
\left(
\begin{array}{c}
\widehat{\theta}_{1,n}^{(b+1)} \medskip \\ 
\widehat{\theta}_{2,a, n}^{(b+1)} 
\end{array}
\right)
= 
\left(
\begin{array}{c}
\widehat{\theta}_{1,n}^{(b)} \medskip \\ 
\widehat{\theta}_{2,a, n}^{(b)} 
\end{array}
\right)
+ \alpha_b \Lambda_{a}^{(b)}
\end{equation*} 
\State Set $\widehat{\theta}_{2,a',n}^{(b+1)} =
\widehat{\theta}_{2,a',n}^{(b)}$ for all $a' \ne a$ 
\EndFor
\State If $\max_{a}\bigg|
C_{n}^{\lambda}\left\lbrace
\widehat{\theta}_{1,n}^{(b+1)},  \widehat{\theta}_{2,a,n}^{(b+1)}
\right\rbrace 
- 
C_{n}^{\lambda}\left\lbrace
\widehat{\theta}_{1,n}^{(b)},  \widehat{\theta}_{2,a,n}^{(b)}
\right\rbrace \bigg| \le \epsilon$ stop.  
\EndFor
\Ensure{$\left(\widehat{\theta}_{1,n}, \widehat{\theta}_{2,n}\right) 
= \left(\widehat{\theta}_{1,n}^{(b+1)},
  \widehat{\theta}_{2,n}^{(b+1)}\right)$}
\end{algorithmic}
\end{algorithm}

To select the dimension of the feature map we choose the lowest
dimension for which the Brownian distance covariance test of
independence between
$\bY^{t+1} - \mathcal{L}_{M_1+M_2}(\bS^t; \widehat{\theta}_{1,n},
\widehat{\theta}_{2,A^t,n})$
and $\bS^{t}$ fails to reject at a pre-specified error level
$\tau \in (0,1)$.  Let $\widehat{\phi}_{n}^1$ be the estimated feature
map $\bs \mapsto \mathcal{L}_{M_1}(\bs;\widehat{\theta}_{1})$.  Define
$\widehat{R}_{n}^1 = \left\lbrace j\in\left\lbrace 1,\ldots
    r_1\right\rbrace\,:\, \widehat{\Sigma}_{1,\ell,j}^2 \neq 0\,\mbox{for
    some}\,\ell\in\{1,\ldots, r_2\}\right\rbrace$
to be the elements of $\bS^t$ that dictate
$\bS_{\widehat{\phi}_{n}^1}^t$; write $\bS_{\widehat{R}_{n}^1}^t$ as
shorthand for
$\left\lbrace S^t_{j}\right\rbrace_{j\in\widehat{R}_{n}^1}$.  One may
wish to iterate the foregoing estimation procedure as described in
Corollary \ref{coraline}.  However, because the components of
$\bS_{\widehat{\phi}_{n}^1}^t$ are each a potentially nonlinear
combination of the elements of $\bS_{\widehat{R}_{n}^1}^t$, therefore
a sparse feature map defined on the domain of
$\bS_{\widehat{\phi}_{n}^{1}}^t$ may not be any more sparse in terms
of the original features.  Thus, when iterating the feature map
construction algorithm, we recommend using the reduced process
$\left\lbrace \overline{\bS}_{\widehat{R}_{n}^1, i}^{T+1},
  \overline{\bA}_{i}^T,
  \overline{\bU}_{i}^{T}\right\rbrace_{i=1}^{n}$ and the input;
because the sigma-algebra generated by $\bS_{\widehat{R}_{n}^1}^t$ contains
the sigma-algebra generated by $\bS_{\widehat{\phi}_{n}^1}^t$, this
does not incur any loss in generality.  The above procedure can be
iterated until no further dimension reduction occurs.

\section{Simulation Experiments}
\label{sec:experiments}
We evaluate the finite sample performance of the proposed
method (pre-screening with Brownian distance covariance + iterative alternating deep neural networks, which we will simply refer to as ADNN in this section) using a series of simulation experiments.  
To form a basis for comparison, 
we consider two alternative feature construction methods:
 (PCA) principal components analysis, so that 
the estimated feature map $\widehat{\phi}_{\mathrm{PCA}}(\bs)$ is the
projection of $\bs$ onto the first $k$ principal components of 
$T^{-1}\sum_{t=1}^{T}\pn \left\lbrace \bS^t - \pn \bS^t\right\rbrace
\left\lbrace \bS^t - \pn \bS^t\right\rbrace^{\T}$; and (tNN) a
traditional sparse neural network, which can be seen as a special case of
our proposed alternating deep neural network estimator where there is only 1 action. In our implementation of PCA, we choose the number of principal components, $k$, corresponding to 90\% of variance explained. We do not compare with sparse PCA for variable selection, because based on preliminary runs, the principal components that explain 50\% of variance already use all the variables in our generative model. In our implementation of tNN, we build a separate tNN for each $a \in \mathcal{A}$, where $(\lambda, K_1, K_2, r_1)$ are tuned using cross-validation, and take the union of selected variables and constructed features. Note that there is no other obvious way to join the constructed features from tNN but to simply concatenate them, which will lead to inefficient dimension reduction especially when $|\mathcal{A}|$ is large, whereas we will see that ADNN provides a much more efficient way to aggregate the useful information across actions.

We evaluate the quality of a feature map, $\phi$, in terms of the
marginal mean outcome under the estimated optimal regime constructed
from the reduced data
$\left\lbrace \overline{\bS}_{i}^{T+1}, \overline{\bA}_{i}^{T},
  \overline{\bU}_{i}^{T}\right\rbrace_{i=1}^{n}$
using Q-learning with function approximation
\citep[][]{Bertsekas1996,murphy2005generalization}; we use both linear
function approximation and non-linear function approximation with neural networks.  A description of Q-learning as well as these
approximation architectures are described in the Supplemental
Materials. 

\begin{figure}[!h]
	\makebox[\textwidth][c]{\includegraphics[width=0.85\textwidth]{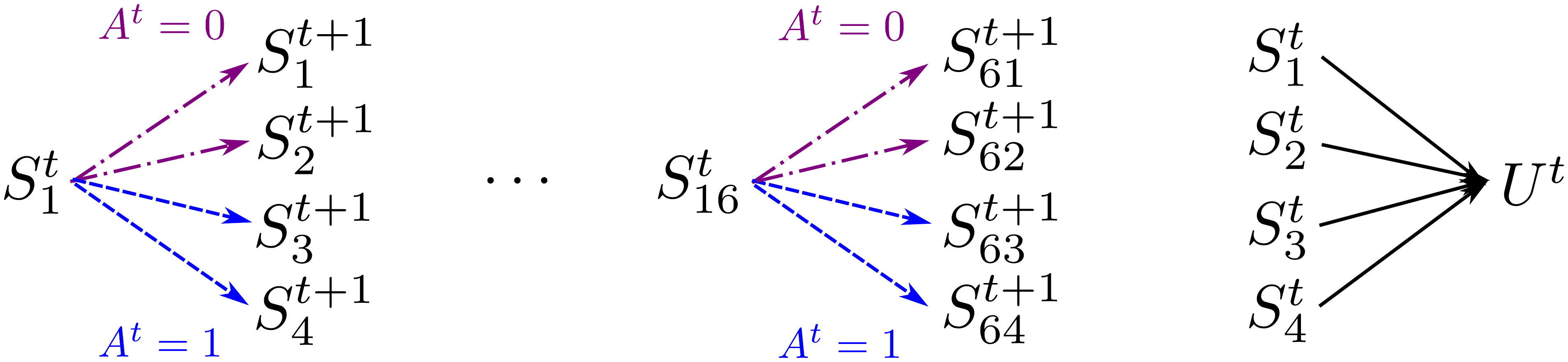}}
	\caption{\label{MDPGen}
          Relationship between $\bS^t$ and $\mathbf{Y}^{t+1}$ in the generative model, which depends on the action. First 16 variables determine the next state. First 4 variables determine the utility.
        }
\end{figure}

We consider data from the following class of generative models, as illustrated in Figure \ref{MDPGen}:
$$
\begin{aligned}
&\bS^1 \sim \mathrm{Normal}_{64}\left(0, \; 0.25\mathbf{I}_{64}\right); \quad
A^1,\ldots, A^T\sim_{i.i.d.} \mathrm{Bernoulli}\left( 0.5 \right); \\ 
&S^{t+1}_{4i-3}, S^{t+1}_{4i-2} \sim_{i.i.d.} \mathrm{Normal}\left\{(1-A^t)g(S_i^t), \; 0.01(1-A^t) + 0.25A^t\right\}; \\
&S^{t+1}_{4i-1}, S^{t+1}_{4i} \sim_{i.i.d.} \mathrm{Normal}\left\{A^tg(S_i^t), \; 0.01A^t + 0.25(1-A^t)\right\}; \\
&U^t \sim \mathrm{Normal}\{(1-A^t)[2\{g(S^t_1) + g(S_2^t)\} - \{g(S_3^t) + g(S_4^t)\}] \\
& \qquad \qquad \qquad+ A^t[2\{g(S^t_3) + g(S_4^t)\} - \{g(S_1^t) + g(S_2^t)\}], \; 0.01\}; \\
&\text{for} \; i = 1, 2, \ldots, 16.
\end{aligned}
$$

The above class of models is indexed by
$g:\mathbb{R}\rightarrow \mathbb{R}$ which we vary across the
following maps: identity $g(u)=u$, truncated quadratic $g(u)=\min\{u^2, 3\}$, and truncated exponential $g(u) = \min\{\exp(u), 3\}$, where the truncation is used to keep all variables of relatively the same scale across time points. Additionally, we add 3 types of noise variables, each taking up about $\frac{1}{3}$ of total noises added: (i) dependent noise variables $D^t_j$, which are generated the same way as above except that they don't affect the utility; (ii) white noises $W^t_k$, which are sampled independently from $\mathrm{Normal}(0, 0.25)$ at each time point; and (iii) constants $C^t_l$, which are sampled independently from $\mathrm{Normal}(0, 0.25)$ at $t=1$ and remain constant over time. More precisely, let $m$ be the total number of noise variables, then
$$
\begin{aligned}
&D^1_j, W^1_k, C^1_l \sim_{i.i.d.} \mathrm{Normal}\left(0, 0.25\right); \\
&D^{t+1}_{4i-3}, D^{t+1}_{4i-2} \sim_{i.i.d.} \mathrm{Normal}\left\{(1-A^t)g(D_i^t), \; 0.01(1-A^t) + 0.25A^t\right\}; \\
&D^{t+1}_{4i-1}, D^{t+1}_{4i} \sim_{i.i.d.} \mathrm{Normal}\left\{A^tg(D_i^t), \; 0.01A^t + 0.25(1-A^t)\right\}; \\
&W^t_k \sim_{i.i.d.} \mathrm{Normal}\left(0, 0.25\right); \; C^t_l = C^1_l; \\
&\text{for} \; j = 1, 2, \ldots, \lfloor m/3 \rfloor; k = 1, 2, \ldots, \lceil m/3 \rceil;  l = 1, 2, \ldots, \lceil m/3 \rceil.
\end{aligned}
$$
It can be seen that the first 16 variables, the first 4 variables, and $\{g(S_1^t), g(S_2^t), g(S_3^t) + g(S_4^t)\}^{\T}$ all induce a sufficient MDP. 
the foregoing class of models is designed to evaluate the ability of
the proposed method to identify low-dimensional and potentially
nonlinear features of the state in the presence of action-dependent transitions and various noises. 
For each Monte Carlo replication, we sample $n=30$ i.i.d. trajectories 
of length $T=90$ from the above generative model.   

The results based on 500 Monte Carlo replications 
are reported in Table \ref{resultsSim1} - \ref{resultsSim3}.  In addition 
to reporting the marginal mean outcome under the policy estimated
using Q-learning with both function approximations, we also report: (nVar) the number of selected variables; and (nDim) the dimension of the feature map. The table shows that (i) ADNN produces substantially smaller nVar and nDim compared with PCA or tNN in all cases; (ii) ADNN is robust to the 3 types of noises; (iii) when fed into the Q-learning algorithm, ADNN leads to considerably better marginal mean outcome than PCA and the original states under non-linear models; and (iv) ADNN is able to construct features suitable for Q-learning with linear function approximation even when the utility function and transition between states are non-linear.

\begin{table}[h!]
	\begin{center}
		\begin{tabular}{c|c|c|c|c|c|c}
			\hline
			\multirow{2}{*}{Model} & \multirow{2}{*}{nNoise}& \multirow{2}{*}{Feature map} & \multirow{2}{*}{Linear Q} 
                   & \multirow{2}{*}{NNQ} & \multirow{2}{*}{nVar} & \multirow{2}{*}{nDim} \\ 
                   &&&&&& \\ 
			\hline
                   \multirow{15}{*}{linear} & \multirow{5}{*}{0} & $\bs_t$ & 3.36(0.012) & 3.31(0.012) & 64 & 64 \\
                   & & $(s_1^t, s_2^t, s_3^t, s_4^t)^\T$ & 3.34(0.012) & 3.31(0.013) & 4 & 4 \\
                   & & $\widehat{\phi}_\text{ADNN}(\bs^t)$ & 3.21(0.018) & 3.34(0.013) & 4.1(0.01) & 3.1(0.02) \\
                   & & $\widehat{\phi}_\text{tNN}(\bs^t)$ & 3.38(0.012) & 3.30(0.012) & 16.0(0.00) & 34.4(0.13) \\
                   & & $\widehat{\phi}_\text{PCA}(\bs^t)$ & 3.34(0.012) & 3.30(0.012) & 64 & 50.0(0.00) \\
			\cline{2-7}
                   & \multirow{5}{*}{50} & $\bs_t$ & 3.31(0.012) & 3.27(0.012) & 114 & 114 \\
                   & & $(s_1^t, s_2^t, s_3^t, s_4^t)^\T$ & 3.31(0.012) & 3.29(0.013) & 4 & 4 \\
                   & & $\widehat{\phi}_\text{ADNN}(\bs^t)$ & 3.26(0.014) & 3.32(0.013) & 5.6(0.08) & 4.6(0.08) \\
                   & & $\widehat{\phi}_\text{tNN}(\bs^t)$ & 3.32(0.012) & 3.29(0.013) & 37.0(0.00) & 86.0(0.19) \\
                   & & $\widehat{\phi}_\text{PCA}(\bs^t)$ & 3.34(0.012) & 3.28(0.013) & 114 & 85.8(0.02) \\
			\cline{2-7}
                   & \multirow{5}{*}{200} & $\bs_t$ & 2.17(0.016) & 2.98(0.035) & 264 & 264 \\
                   & & $(s_1^t, s_2^t, s_3^t, s_4^t)^\T$ & 3.33(0.012) & 3.31(0.013) & 4 & 4 \\
                   & & $\widehat{\phi}_\text{ADNN}(\bs^t)$ & 3.29(0.013) & 3.32(0.012) & 10.2(0.12) & 7.5(0.09) \\
                   & & $\widehat{\phi}_\text{tNN}(\bs^t)$ & 3.34(0.012) & 3.27(0.013) & 87.4(0.11) & 157.8(0.48) \\
                   & & $\widehat{\phi}_\text{PCA}(\bs^t)$ & 3.33(0.013) & 3.11(0.028) & 264 & 166.0(0.02) \\
			\hline
		\end{tabular}
                \caption{\label{resultsSim1}
                  Comparison of feature map estimators under linear transition and different number of noise variables (nNoise) 
                  in terms of: marginal mean outcome using Q-learning with 
                  linear function approximation (Linear Q); Q-learning with neural network function
                  approximation
                  (NN Q); the number of selected variables (nVar); and the dimension of the feature map (nDim)}
	\end{center}
\end{table}

\begin{table}[h!]
	\begin{center}
		\begin{tabular}{c|c|c|c|c|c|c}
			\hline
			\multirow{2}{*}{Model} & \multirow{2}{*}{nNoise}& \multirow{2}{*}{Feature map} & \multirow{2}{*}{Linear Q} 
                   & \multirow{2}{*}{NNQ} & \multirow{2}{*}{nVar} & \multirow{2}{*}{nDim} \\ 
                   &&&&&& \\ 
			\hline
                   \multirow{15}{*}{quad} & \multirow{5}{*}{0} & $\bs_t$ & 3.08(0.062) & 2.64(0.073) & 64 & 64 \\
                   & & $(s_1^t, s_2^t, s_3^t, s_4^t)^\T$ & 2.54(0.056) & 6.75(0.046) & 4 & 4 \\
                   & & $\widehat{\phi}_\text{ADNN}(\bs^t)$ & 6.63(0.038) & 6.97(0.034) & 4.1(0.02) & 2.4(0.04) \\
                   & & $\widehat{\phi}_\text{tNN}(\bs^t)$ & 6.94(0.027) & 6.54(0.068) & 15.3(0.04) & 37.1(0.22) \\
                   & & $\widehat{\phi}_\text{PCA}(\bs^t)$ & 2.97(0.064) & 2.50(0.067) & 64 & 51.2(0.02) \\
			\cline{2-7}
                   & \multirow{5}{*}{50} & $\bs_t$ & 2.96(0.054) & 1.69(0.064) & 114 & 114 \\
                   & & $(s_1^t, s_2^t, s_3^t, s_4^t)^\T$ & 2.58(0.057) & 6.76(0.042) & 4 & 4 \\
                   & & $\widehat{\phi}_\text{ADNN}(\bs^t)$ & 6.76(0.032) & 6.99(0.030) & 6.4(0.06) & 5.3(0.09) \\
                   & & $\widehat{\phi}_\text{tNN}(\bs^t)$ & 6.98(0.031) & 6.53(0.064) & 36.5(0.03) & 88.3(0.22) \\
                   & & $\widehat{\phi}_\text{PCA}(\bs^t)$ & 3.09(0.061) & 2.00(0.067) & 114 & 87.1(0.03) \\
			\cline{2-7}
                   & \multirow{5}{*}{200} & $\bs_t$ & 1.28(0.030) & 0.88(0.031) & 264 & 264 \\
                   & & $(s_1^t, s_2^t, s_3^t, s_4^t)^\T$ & 2.52(0.056) & 6.68(0.050) & 4 & 4 \\
                   & & $\widehat{\phi}_\text{ADNN}(\bs^t)$ & 6.87(0.034) & 6.92(0.033) & 14.3(0.14) & 12.5(0.23) \\
                   & & $\widehat{\phi}_\text{tNN}(\bs^t)$ & 6.76(0.044) & 6.03(0.075) & 84.1(0.11) & 152.4(0.36) \\
                   & & $\widehat{\phi}_\text{PCA}(\bs^t)$ & 3.09(0.062) & 0.96(0.033) & 264 & 167.4(0.03) \\
			\hline
		\end{tabular}
                \caption{\label{resultsSim2}
                  Comparison of feature map estimators under quadratic transition}
	\end{center}
\end{table}

\begin{table}[h!]
	\begin{center}
		\begin{tabular}{c|c|c|c|c|c|c}
			\hline
			\multirow{2}{*}{Model} & \multirow{2}{*}{nNoise}& \multirow{2}{*}{Feature map} & \multirow{2}{*}{Linear Q} 
                   & \multirow{2}{*}{NNQ} & \multirow{2}{*}{nVar} & \multirow{2}{*}{nDim} \\ 
                   &&&&&& \\ 
			\hline
                   \multirow{15}{*}{exp} & \multirow{5}{*}{0} & $\bs_t$ & 8.73(0.008) & 8.78(0.012) & 64 & 64 \\
                   & & $(s_1^t, s_2^t, s_3^t, s_4^t)^\T$ & 9.20(0.006) & 9.43(0.004) & 4 & 4 \\
                   & & $\widehat{\phi}_\text{ADNN}(\bs^t)$ & 9.30(0.018) & 9.45(0.004) & 4.3(0.13) & 2.4(0.13) \\
                   & & $\widehat{\phi}_\text{tNN}(\bs^t)$ & 9.44(0.005) & 9.29(0.009) & 16.0(0.00) & 42.3(0.17) \\
                   & & $\widehat{\phi}_\text{PCA}(\bs^t)$ & 9.10(0.016) & 9.02(0.023) & 64 & 14.2(0.018) \\
			\cline{2-7}
                   & \multirow{5}{*}{50} & $\bs_t$ & 8.78(0.008) & 8.77(0.012) & 114 & 114 \\
                   & & $(s_1^t, s_2^t, s_3^t, s_4^t)^\T$ & 9.19(0.006) & 9.43(0.005) & 4 & 4 \\
                   & & $\widehat{\phi}_\text{ADNN}(\bs^t)$ & 9.32(0.018) & 9.43(0.005) & 5.4(0.03) & 2.4(0.04) \\
                   & & $\widehat{\phi}_\text{tNN}(\bs^t)$ & 9.43(0.005) & 9.18(0.012) & 37.0(0.00) & 81.5(0.28) \\
                   & & $\widehat{\phi}_\text{PCA}(\bs^t)$ & 8.89(0.014) & 8.99(0.020) & 114 & 37.2(0.02) \\
			\cline{2-7}
                   & \multirow{5}{*}{200} & $\bs_t$ & 8.71(0.008) & 8.73(0.012) & 264 & 264 \\
                   & & $(s_1^t, s_2^t, s_3^t, s_4^t)^\T$ & 9.19(0.006) & 9.44(0.004) & 4 & 4 \\
                   & & $\widehat{\phi}_\text{ADNN}(\bs^t)$ & 9.37(0.016) & 9.41(0.006) & 7.8(0.09) & 3.3(0.12) \\
                   & & $\widehat{\phi}_\text{tNN}(\bs^t)$ & 9.41(0.005) & 9.06(0.016) & 93.4(0.10) & 152.4(0.42) \\
                   & & $\widehat{\phi}_\text{PCA}(\bs^t)$ & 8.66(0.014) & 9.02(0.022) & 264 & 91.4(0.02) \\
			\hline
		\end{tabular}
                \caption{\label{resultsSim3}
                  Comparison of feature map estimators under exponential transition}
	\end{center}
\end{table}

\section{Application to BASICS-Mobile}
\label{sec:mobileHealth}
We illustrate the proposed methedology using data on the effectiveness
of BASICS-Mobile, a behavioral intervention delivered via mobile
device, targeting heavy drinking and smoking among college students
\citep[][]{Witkiewitz2014}.    
Mobile interventions are appealing because of their 24-hour
availability, anonymity, portability, increased compliance, and
accurate data recording \citep[][]{Heron2010}. BASICS-Mobile enrolled
30 students and lasted for 14 days. On the afternoon and evening of
each day, the student is asked to complete a list of self-report
questions, and then either an informational module or a treatment
module is provided. A treatment module contains 1-3 mobile phone
screens of interactive content, such as comparing the student's
smoking level with the levels of their peers, or guiding the student to
manage their smoking urges.
A treatment module is generally more burdensome than an informational
module, and may be less effective if, for example, the student's stress
level is high; furthermore, excessive treatment can cause
habituation and disengagement from the intervention.   
 An optimal intervention will assignment a treatment
module if and when it is needed without diminishing 
engagement.   

In our original formulation of this decision problem as an MDP, the 
state comprises of 15 variables capturing 
baseline information, current answers to the self-report
questions,  a weekend indicator, age, past attempts to quit
smoking, current smoking urge, and current stress level; the
action is whether a treatment module gets assigned; the reward is the
negative of the cigarettes smoked at the next time point; the goal is
to find a strategy that minimizes cumulative cigarette rate.

\begin{figure}[!h]
	\makebox[\textwidth][c]{\includegraphics[width=1.05\textwidth]{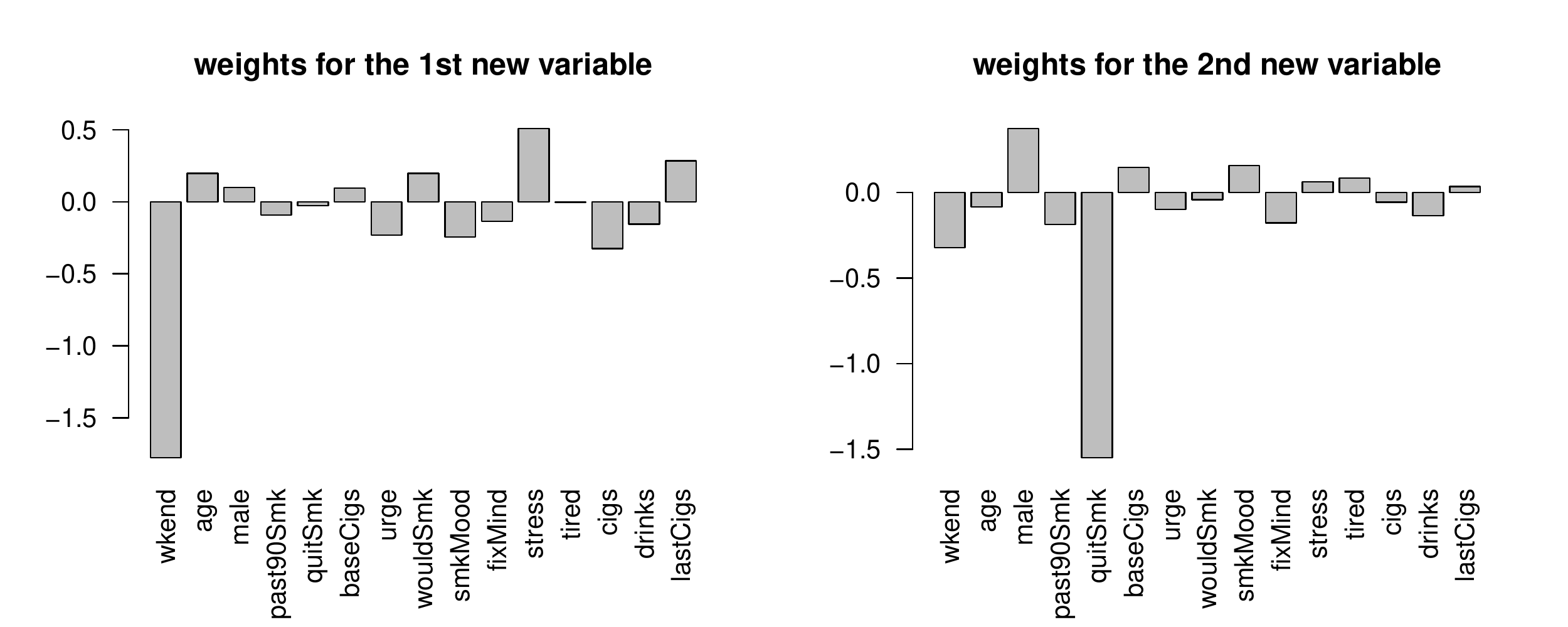}}
	\caption{Weights of the original variables in the first two
          components 
          of the estimated feature map.}
	\label{newVar}
\end{figure}

Two students with large amounts of missing data are excluded. All
other missing values are imputed with the fitted value from a local
polynomial regression of the state variable on time $t$. We treat
all the variables as ordinal, partitioning some of them (see the
Supplemental
Materials for a complete description). We estimate
$\widehat{\phi}_\text{ADNN}(s_t)$ wherein conditional
independence is checked via condition (i) in Lemma 3.4.  
The dimension of
$\widehat{\phi}_\text{ADNN}(s_t)$ is set to be the smallest dimension
for which $\widehat{\phi}_\text{ADNN}(s_t)$ fails to reject this
independence
condition at level $\tau = 0.05$; this procedure resulted in a 
feature of dimension six.  
To increase the interpretability of the constructed
feature map, we constrained the dimension reduction network to have no
hidden layers. Under this
constraint, $\widehat{\phi}_\text{ADNN}(s_t)$ is a linear
transformation of $s_t$ followed by application of  $\Phi^{\circ}$
which was set to be the arctangent function. A plot of the
weights of the 15 original variables in the linear transformation for
each component of the feature map is useful in interpreting the
learned feature map; see Figure \ref{newVar} for an example.  

We estimate the optimal strategy using Q-learning applied to the
learned feature map.  
Comparing the estimated parameters for treatment and no treatment,
while examining the plots of weights, we can give a sense of how the
original variables impact the optimal treatment assignment. For
instance, the 1st parameter in the Q-function for treatment is smaller
than the one for no treatment, which suggests that the 1st new variable
contributes to the decision to apply treatment by being small, i.e.,
if it is the weekend and a student's stress level is low then the
estimated
policy is more likely to provide treatment.  
This agrees with the intuition that a treatment
module would be more effective when the student is not busy or
stressed. Similarly, it can be seen that previous attempts to quit smoking is
positively
associated with providing treatment, with the possible explanation that individuals with prior quit attempts tend to be more severe and in need of frequent treatment.

\section{Discussion}
\label{sec:discussion}
Data-driven decision support systems are being deployed across a wide
range of application domains including medicine, engineering, and
business.  MDPs provide the mathematical underpinning for most
data-driven decision problems with an infinite or indefinite time
horizon. While the MDP model is extremely general, choosing 
a parsimonious representation of a decision process that fits the MDP 
model is non-trivial.  We introduced the notion of a feature map which 
induces a sufficient MDP and provided an estimator of such a feature
map based on a variant of deep neural networks.

There are several important ways in which this work can be extended;
we mention two of the most pressing here.  We
considered estimation from a batch of i.i.d.  replicates; however, in
some applications it may be desirable to estimate a feature map online
as data accumulate.  In such cases, a data-driven, and hence evolving,
feature map of the state will be stored complicating estimation.
Furthermore, because the proposed algorithm sweeps through the
observed data multiple times it is not suitable for real-time
estimation.  Another important extension is to states with complex
data structures, e.g., images and text, such data are increasingly
common in health, engineering, and security applications.  Existing
neural network architectures designed for such data
\citep[][]{krizhevsky2012imagenet, dahl2012context, simonyan2014very}
could potentially be integrated into the proposed feature map
construction algorithm.

\nocite{*}
\bibliographystyle{jasa}
\bibliography{myBib}

\section{Supplemental Materials}

\noindent \textbf{Q-learning} 

Let $(\overline{\mathbf{A}}^{t}, \overline{\mathbf{S}}^{t+1}, \overline{\mathbf{U}}^t)$ be an MDP. The value of a state-action pair under a policy $\pi$, referred to as the Q-function, is the discounted mean utility if the current state is $\mathbf{s}$, current action is $a$, and the agent follows $\pi$ afterwards: $Q^{\pi}(\mathbf{s}, a) = \mathbb{E}\left\{\sum_{t \geq 1} \gamma^{t-1} U^{*t}(\pi) \lvert \mathbf{S}^1=\mathbf{s}, A^1=a \right\}$, where $\gamma \in (0, 1)$ is the discount factor. An optimal policy $\pi^\text{opt}$ yields the largest value for every state-action pair. Denote the corresponding Q-function as $Q^\text{opt}(\mathbf{s}, a) = Q^{\pi^\text{opt}}(\mathbf{s}, a) = \underset{\pi}{\max} \; Q^\pi(\mathbf{s}, a)$. If $Q^\text{opt}(\mathbf{s}, a)$ is known for all $(\mathbf{s}, a)$, then an optimal policy can be defined as $\pi^\text{opt}(\bs) = \underset{a \in \mathcal{A}}{\text{argmax}} \; Q^\text{opt}(\mathbf{s}, a).$ $Q^\text{opt}$ satisfies the Bellman Optimality Equations (BOE): 
$$
Q^\text{opt}(\mathbf{s}, a) = \mathbb{E}\left\{U^t + \gamma \; \underset{a' \in \mathcal{A}}{\max} \; Q^\text{opt}(\mathbf{S}^{t+1}, a') \; \lvert \; \mathbf{S}^t=\bs, A^t=a\right\}.
$$
In practice, one often cannot obtain $Q^\text{opt}$ by solving the above equations, because computing the right-hand-side requires the underlying model of the MDP, which is often unknown. Besides, solving a huge linear system can be costly. 

Q-learning is a stochastic optimization algorithm that doesn't require knowing the transition model or solving a linear system. The update step for the classic Q-learning, Watkin's Q-learning \citep[][]{sutton1998}, for finite state and action spaces, is as follows:
$$
Q^{k+1}(\mathbf{s}_t, a_t) \leftarrow Q^k(\mathbf{s}_t, a_t) + \alpha \{u_t + \gamma \, \underset{a}{\max} \; Q^k(\mathbf{s}_{t+1}, a) - Q^k(\mathbf{s}_t, a_t)\},
$$
where $\alpha$ is the learning rate. If the state space is continuous, one may approximate $Q(\mathbf{s}, a)$ with a parametric function $F(\mathbf{s}, a; \boldsymbol{\theta})$ and update the parameters instead:
$$
\boldsymbol{\theta}^{k+1} \leftarrow \boldsymbol{\theta}^k + \alpha \{u_t + \gamma \; \underset{a}{\max} \; F(\mathbf{s}_{t+1}, a; \boldsymbol{\theta}^k) - F(\mathbf{s}_t, a_t; \boldsymbol{\theta}^k)\} \cdot \nabla_{\! \boldsymbol{\theta}} F(\mathbf{s}_t, a_t; \boldsymbol{\theta}^k).
$$

\noindent \textbf{Proof of Theorem \ref{theoremey}}
\begin{proof}
First we show that the process $(\overline{\mathbf{A}}^{t}, \overline{\mathbf{S}}^{t+1}_{\phi}, \overline{\mathbf{U}}^t)$ induced by $(\phi, \Pi_{\phi, msbl})$ satisfies (SM1). For any $t \in \mathbb{N}$ and measurable subset $\mathcal{G} \in \mathbb{R}^q$,
$$
\begin{aligned}
&P(\mathbf{S}_\phi^{t+1} \in \mathcal{G}_\phi^{t+1} \lvert \overline{\mathbf{S}}_\phi^t, \overline{\bA}^t) \\
= \, &\mathbb{E}\{P(\mathbf{S}_\phi^{t+1} \in \mathcal{G}_\phi^{t+1} \lvert \mathbf{S}^t, \overline{\mathbf{S}}_\phi^t, \overline{\bA}^t) \lvert \overline{\mathbf{S}}_\phi^t, \overline{\bA}^t\} \\
= \, &\mathbb{E}\{P(\mathbf{S}_\phi^{t+1} \in \mathcal{G}_\phi^{t+1} \lvert \mathbf{S}^t, \mathbf{S}_\phi^t, A^t) \lvert \overline{\mathbf{S}}_\phi^t, \overline{\bA}^t\} \quad (\text{by Markov property of the original process}) \\
= \, &\mathbb{E}\{P(\mathbf{S}_\phi^{t+1} \in \mathcal{G}_\phi^{t+1} \lvert \mathbf{S}_\phi^t, A^t) \lvert \overline{\mathbf{S}}_\phi^t, \overline{\bA}^t\} \quad (\text{by (\ref{sufficientTheoremCond1})}) \\
= \, &P(\mathbf{S}^{t+1}_\phi \in \mathcal{G}_\phi^{t+1} \lvert \mathbf{S}^t_\phi, A^t)
\end{aligned}
$$
Also note that $\mathbb{E}\{P(\bS_\phi^{t+1} \in \mathcal{G}_\phi^{t+1} \lvert \mathbf{S}^t, \overline{\bS}_\phi^t, \overline{\bA}^t) \lvert \overline{\bS}_\phi^t, \overline{\bA}^t\}$ does not depend on $t$ by homogeneity of the original process. Thus the induced process is Markov and homogeneous.

Next we show that the induced process satisfies (SM2). Let $Q^\text{opt}(\mathbf{s}, a)$ be defined as before. Then we have
$$
\begin{aligned}
Q^\text{opt}(\mathbf{s}, a) &= \text{E}[U^t + \gamma \; \underset{a'}{\max} \; Q^\text{opt}(\mathbf{S}^{t+1}, a') \; \lvert \; \mathbf{S}^t=\mathbf{s}, A^t=a] \quad (\text{by BOE})\\
&= \text{E}[U^t + \gamma \; \underset{a'}{\max} \; Q^\text{opt}(\mathbf{S}^{t+1}, a') \; \lvert \; \mathbf{S}^t=\mathbf{s}, \mathbf{S}^t_\phi=\mathbf{s}_\phi, A^t=a] \\
&= \text{E}[U^t + \gamma \; \underset{a'}{\max} \; Q^\text{opt}(\mathbf{S}^{t+1}, a') \; \lvert \; \mathbf{S}^t_\phi=\mathbf{s}_\phi, A^t=a] \quad (\text{by (2)}) \\
&= Q_\phi^\text{opt}(\mathbf{s}_\phi, a), \quad \text{and}
\end{aligned}
$$
$$
\pi^\text{opt}(\mathbf{s}) = \underset{a}{\text{argmax}} \; Q^\text{opt}(\mathbf{s}, a) = \underset{a}{\text{argmax}} \; Q_\phi^\text{opt}(\mathbf{s}_\phi, a) = \pi_\phi^\text{opt}(\mathbf{s}_\phi).
$$
\end{proof}

\noindent \textbf{Proof of Corollary \ref{coraline}}
\begin{proof}
By assumption $(\phi_0, \Pi_{\phi_0, \text{msrbl}})$ induces a sufficient MDP for $\pi^\text{opt}$ within $\Pi_\text{msrbl}$, then by definition the process $(\overline{\mathbf{A}}^t, \overline{\mathbf{S}}^{t+1}_{\phi_0}, \overline{\mathbf{U}}^t)$ is Markov and homogeneous, and there exists $\pi^\text{opt} = \pi_{\phi_0}^\text{opt} \circ \phi_0$.

Define $\overline{\phi}_k = \phi_k \circ \cdots \circ \phi_0$. By (3) and Theorem 3.2, $(\phi_1, \Pi_{\phi_1,\text{msrbl}})$ induces a sufficient MDP for $\pi_{\phi_0}^\text{opt}$ within $\Pi_{\phi_0, \text{msrbl}}$. Then the process $(\overline{\mathbf{A}}^{t}, \overline{\mathbf{S}}_{\overline{\phi}_1}, \overline{\mathbf{U}}^t)$ is Markov and homogeneous, and there exists $\pi^\text{opt}_{\phi_0} = \pi^\text{opt}_{\overline{\phi}_1} \circ \phi_1$. Thus $\pi^\text{opt} = \pi_{\phi_0}^\text{opt} \circ \phi_0 = \pi^\text{opt}_{\overline{\phi}_1} \circ \overline{\phi}_1$. Therefore $(\overline{\phi}_1, \Pi_{\overline{\phi}_1,\text{msrbl}})$ induces a sufficient MDP for $\pi^\text{opt}$ within $\Pi_\text{msrbl}.$
\end{proof}

\noindent \textbf{Proof of Lemma \ref{llama}}
\begin{proof}
We show that (i) $\Rightarrow \mathbf{Y}^{t+1} \independent \mathbf{S}^t \lvert \mathbf{S}_\phi^t, A^t$:
$$
\begin{aligned}
&\{\mathbf{Y}^{t+1} - \mathbb{E}(\bY^{t+1}\lvert\mathbf{S}_\phi^t, A^t)\} \independent \mathbf{S}^t \lvert A^t \\
\Rightarrow &\{\mathbf{Y}^{t+1} - \mathbb{E}(\bY^{t+1}\lvert\mathbf{S}_\phi^t, A^t)\} \independent (\mathbf{S}^t, \mathbf{S}_\phi^t) \lvert A_t \\
\Rightarrow &\{\mathbf{Y}^{t+1} - \mathbb{E}(\bY^{t+1}\lvert\mathbf{S}_\phi^t, A^t)\} \independent \mathbf{S}^t \lvert \mathbf{S}_\phi^t, A^t \\
\Rightarrow &\mathbf{Y}^{t+1} \independent \mathbf{S}^t \lvert \mathbf{S}_\phi^t, A^t.
\end{aligned}
$$
The 1st implication follows from the fact that $\bS_\phi^t$ is a transformation of $\bS^t$. The 2nd implication follows from the fact that $X \independent Y, Z \Rightarrow f(X \lvert Y, Z) = f(X) = f(X \lvert Z) \Rightarrow X \independent Y \lvert Z$ for random variables $X, Y,$ and $Z$. The 3rd implication follows from the fact that $\mathbb{E}(\bY^{t+1}\lvert\mathbf{S}_\phi^t, A^t)$ is constant conditional on $\bS^t_\phi$ and $A^t$.

Similarly, one can show that (ii) $\Rightarrow \mathbf{Y}^{t+1} \independent \mathbf{S}^t \lvert \mathbf{S}_\phi^t, A^t$.
\end{proof}

\noindent \textbf{Proof of Theorem \ref{strongThm}}
\begin{proof}
First we show that the process $(\overline{\mathbf{A}}^{t}, \overline{\mathbf{S}}^{t+1}_{\phi}, \overline{\mathbf{U}}^t)$ induced by $(\phi, \Pi_{\phi, msbl})$ satisfies (SM1). For any $t \in \mathbb{N}$ and measurable subset $\mathcal{G} \in \mathbb{R}^q$,
$$
\begin{aligned}
&P(\mathbf{S}_\phi^{t+1} \in \mathcal{G}_\phi^{t+1} \lvert \overline{\mathbf{S}}_\phi^t, \overline{\bA}^t) \\
= \, &\mathbb{E}\{P(\mathbf{S}_\phi^{t+1} \in \mathcal{G}_\phi^{t+1} \lvert \mathbf{S}^t, \overline{\mathbf{S}}_\phi^t, \overline{\bA}^t) \lvert \overline{\mathbf{S}}_\phi^t, \overline{\bA}^t\} \\
= \, &\mathbb{E}\{P(\mathbf{S}_\phi^{t+1} \in \mathcal{G}_\phi^{t+1} \lvert \mathbf{S}^t, \mathbf{S}_\phi^t, A^t) \lvert \overline{\mathbf{S}}_\phi^t, \overline{\bA}^t\} \quad (\text{by Markov property of the original process}) \\
= \, &\mathbb{E}\{P(\mathbf{S}_\phi^{t+1} \in \mathcal{G}_\phi^{t+1} \lvert \mathbf{S}_\phi^t, A^t) \lvert \overline{\mathbf{S}}_\phi^t, \overline{\bA}^t\} \quad (\text{by (\ref{sufficientTheoremCond2})}) \\
= \, &P(\mathbf{S}^{t+1}_\phi \in \mathcal{G}_\phi^{t+1} \lvert \mathbf{S}^t_\phi, A^t)
\end{aligned}
$$
Also note that $\mathbb{E}\{P(\bS_\phi^{t+1} \in \mathcal{G}_\phi^{t+1} \lvert \mathbf{S}^t, \overline{\bS}_\phi^t, \overline{\bA}^t) \lvert \overline{\bS}_\phi^t, \overline{\bA}^t\}$ does not depend on $t$ by homogeneity of the original process. Thus the induced process is Markov and homogeneous.

Next we show that the induced process satisfies (SM2). Define
$$
\begin{aligned}
Q^{\text{opt}, 1}(\bs^t, a^t) := &\mathbb{E}\{U(\bS^t, A^t, \mathbf{S}^{t+1}) \; \lvert \; \mathbf{S}^t=\mathbf{s}^t, A^t=a^t\} \\
= \, &\mathbb{E}\{U(\bS_\phi^t, A^t, \bS_\phi^{t+1}) \; \lvert \; \bS^t=\bs^t, A^t=a^t\} \quad \text{(by (\ref{sufficientTheoremCond2}))} \\
= \, &\mathbb{E}\{U(\bS_\phi^t, A^t, \bS_\phi^{t+1}) \; \lvert \; \bS_\phi^t=\bs_\phi^t, A^t=a^t\} \\
= \, &Q^{\text{opt}, 1}_\phi(\bs^t_\phi, a^t)
\end{aligned}
$$
For $T \geq 2$, define
$$
\begin{aligned}
Q^{\text{opt}, T}(\bs^t, a^t) := \, &\mathbb{E}\{U(\bS^t, A^t, \bS^{t+1}) + \gamma \; \underset{a'}{\max} \; Q^{\text{opt}, T-1}(\mathbf{\bS}^{t+1}, a') \; \lvert \; \bS^t=\bs^t, A^t=a^t\} \\
= \, &\mathbb{E}\{U(\bS^t, A^t, \bS^{t+1}) + \gamma \; \underset{a'}{\max} \; Q_\phi^{\text{opt}, T-1}(\mathbf{S}_\phi^{t+1}, a') \; \lvert \; \bS^t=\bs^t, A^t=a^t\} \quad \text{(by induction)} \\
= \, &\mathbb{E}\{U(\bS_\phi^t, A^t, \bS_\phi^{t+1}) + \gamma \; \underset{a'}{\max} \; Q_\phi^{\text{opt}, T-1}(\mathbf{S}_\phi^{t+1}, a') \; \lvert \; \bS_\phi^t=\bs_\phi^t, A^t=a^t\} \quad \text{(by (\ref{sufficientTheoremCond2}))} \\
= \, &Q^{\text{opt}, T}_\phi(\bs^t_\phi, a^t)
\end{aligned}
$$
From now on we use $U^t = U(\bS^t, A^t, \bS^{t+1}) = U(\bS_\phi^t, A^t, \bS_\phi^{t+1})$ for short.

Claim: $\sup_{a^t, \bs^t} \; \lvert Q^{\text{opt}, T}(\bs^t, a^t) - Q^{\text{opt}}(\bs^t, a^t) \lvert = \mathcal{O}(\gamma^T)$.

Given that the utilities are bounded, we have $\sup_{\bs^t, a^t, \bs^{t+1}} \; \lvert u^t \lvert \leq C_1$, and consequently, \\$\sup_{\bs^t, a^t} \; \lvert Q^{\text{opt}}(\bs^t, a^t) \lvert \leq C_2$, for some constants $C_1$ and $C_2$.
$$
\begin{aligned}
&\underset{a^t, \bs^t}{\sup} \; \left\lvert Q^{\text{opt}, 1}(\bs^t, a^t) - Q^{\text{opt}}(\bs^t, a^t) \right\lvert \\
= \, &\underset{a^t, \bs^t}{\sup} \; \left\lvert \mathbb{E}\{U^t \; \lvert \; \bS^t=\bs^t, A^t=a^t\} - \mathbb{E}\{U^t + \gamma \; \underset{a'}{\max}\; Q^\text{opt}(\bS^{t+1}, a') \; \lvert \; \bS^t=\bs^t, A^t=a^t\} \right\lvert \\
= \, &\underset{a^t, \bs^t}{\sup} \; \gamma \, \left\lvert \mathbb{E}\{\underset{a'}{\max}\; Q^\text{opt}(\bS^{t+1}, a') \; \lvert \; \bS^t=\bs^t, A^t=a^t\} \right\lvert \\
\leq \, &\gamma \, C_2 = \mathcal{O}(\gamma).
\end{aligned}
$$
For $T \geq 2$, assume that $\sup_{a^t, \bs^t} \; \lvert Q^{\text{opt}, T-1}(\bs^t, a^t) - Q^{\text{opt}}(\bs^t, a^t) \lvert \leq \gamma^{T-1} \, C_2,$ then 
$$
\sup_{\bs^t} \; \left\lvert \max_{a'} Q^{\text{opt}, T-1}(\bs^t, a') - \max_{a'} Q^{\text{opt}}(\bs^t, a') \right\lvert \leq \gamma^{T-1} \, C_2,
$$ and
$$
\begin{aligned}
&\underset{a^t, \bs^t}{\sup} \; \left\lvert Q^{\text{opt}, T}(\bs^t, a^t) - Q^{\text{opt}}(\bs^t, a^t) \right\lvert \\
= \, &\underset{a^t, \bs^t}{\sup} \;\left\lvert \mathbb{E}\{U^t + \gamma \; \underset{a'}{\max}\; Q^{\text{opt}, T-1}(\bS^{t+1}, a') - U^t - \gamma \; \underset{a'}{\max}\; Q^\text{opt}(\bS^{t+1}, a') \; \lvert \; \bS^t=\bs^t, A^t=a^t\} \right\lvert \\
= \, &\underset{a^t, \bs^t}{\sup} \; \gamma \, \left\lvert\mathbb{E}\{\underset{a'}{\max}\; Q^{\text{opt}, T-1}(\bS^{t+1}, a') - \underset{a'}{\max}\; Q^\text{opt}(\bS^{t+1}, a') \; \lvert \; \bS^t=\bs^t, A^t=a^t\} \right\lvert \\
\leq \, &\gamma \, (\gamma^{T-1} \, C_2) = \mathcal{O}(\gamma^T), \quad \text{(by induction)}
\end{aligned}
$$
which proves the claim. Therefore, $\lim_{T \rightarrow \infty} Q^{\text{opt}, T} (\bs, a) = Q^{\text{opt}} (\bs, a)$ for all $\bs$ and $a$. Similarly, $\lim_{T \rightarrow \infty} Q_\phi^{\text{opt}, T} (\bs_\phi, a) = Q_\phi^{\text{opt}} (\bs_\phi, a)$ for all $\bs_\phi$ and $a$. And we have
$$
\begin{aligned}
\pi^\text{opt}(\mathbf{s}) &= \underset{a}{\text{argmax}} \; Q^\text{opt}(\mathbf{s}, a) = \underset{a}{\text{argmax}} \lim_{T \rightarrow \infty} Q^{\text{opt}, T} (\bs, a) \\
&= \underset{a}{\text{argmax}} \lim_{T \rightarrow \infty} Q_\phi^{\text{opt}, T} (\bs_\phi, a) = \underset{a}{\text{argmax}} \; Q_\phi^{\text{opt}} (\bs_\phi, a) = \pi_\phi^\text{opt}(\mathbf{s}_\phi).
\end{aligned}
$$
\end{proof}

\noindent \textbf{Proof of Theorem \ref{screenThm}}
\begin{proof}
Let $D$ be the set of all indices. Under the assumption that joint dependence implies marginal dependence, by construction $\mathbf{Y}^{t+1}_{J_{K-1}} \independent \bS^t_{D \setminus J_K} | A^t$. Thus $\mathbf{Y}^{t+1}_{J_{K-1}} \independent \bS^t | \bS^t_{J_K}, A^t$. Because $J_{K-1} = J_K$, the result follows. 
\end{proof}

\noindent \textbf{How the variables from BASICS-Mobile are partitioned}

The variable that records the the number of cigarettes smoked in between reports, CIGS, ranges from 0 to 20. The values imputed by local polynomial regression have decimals and are rounded to the nearest integers. CIGS has 18 unique values after rounding, which is still the most among all variables. We divide CIGS by 20 to rescale it to $[0, 1]$, and the rescaled unique values of CIGS will be used as the levels for all variables. All other variables are rescaled to $[0, 1]$ and rounded to the nearest level.

\end{document}